\documentclass[table]{spie}  

\usepackage[nointegrals]{wasysym}  
\usepackage{amsmath,amsfonts,amssymb}
\usepackage{graphicx}
\usepackage[colorlinks=true, allcolors=blue]{hyperref}
\usepackage{tikz}
\usepackage{orcidlink} 

\usepackage{array}
\usepackage{xcolor}
\usepackage{adjustbox}
\usepackage{multirow}

\usepackage{caption}
\usepackage{subcaption}

\usepackage{mathtools}
\usepackage{scalerel}

\usepackage{csquotes}

\title{Flight mask designs of the Roman Space Telescope Coronagraph Instrument}

\author[a,*]{A~J~Eldorado~Riggs\orcidlink{0000-0002-0863-6228}} 
\author[a]{Dwight Moody}
\author[b]{Jessica Gersh-Range}
\author[c]{Dan Sirbu}
\author[c]{Ruslan Belikov}
\author[a]{Eduardo Bendek\orcidlink{0000-0002-9408-8925}}
\author[a]{Vanessa P. Bailey\orcidlink{0000-0002-5407-2806}}
\author[a]{Kunjithapatham Balasubramanian}
\author[a]{Daniel~W.~Wilson}
\author[a]{Scott~A.~Basinger}
\author[d]{John Debes}
\author[e]{Tyler D. Groff\orcidlink{0000-0001-5978-3247}}
\author[f]{N.~Jeremy Kasdin}
\author[a]{Bertrand Mennesson}
\author[a]{Douglas M. Moore}
\author[a]{Garreth~Ruane\orcidlink{0000-0003-4769-1665}}
\author[a]{Erkin Sidick}
\author[a]{Nicholas Siegler}
\author[a]{John Trauger}
\author[d]{Neil T. Zimmerman}

\affil[a]{Jet Propulsion Laboratory, California Institute of Technology, 4800 Oak Grove Dr., Pasadena, CA 91109, USA}
\affil[b]{Princeton University, Department of Mechanical and Aerospace Engineering, Olden Street, Princeton, NJ 08544, USA}
\affil[c]{NASA Ames Research Center, Moffett Field, Mountain View, CA 94035, USA}
\affil[d]{Space Telescope Science Institute, 3700 San Martin Dr, Baltimore, MD 21218, USA}
\affil[e]{NASA Goddard Space Flight Center, 8800 Greenbelt Rd, Greenbelt, MD 20771, USA }
\affil[f]{University of San Francisco, College of Arts and Sciences, 2130 Fulton St, San Francisco, CA 94117, USA}

\authorinfo{Address all correspondence to A.J. Riggs. Email: aj.riggs@jpl.nasa.gov} 

\begin{document} 
\maketitle

\begin{abstract}
Over the past two decades, thousands of confirmed exoplanets have been detected; the next major challenge is to characterize these other worlds and their stellar systems. Much information on the composition and formation of exoplanets and circumstellar debris disks can only be achieved via direct imaging. Direct imaging is challenging because of the small angular separations ($<1$ arcsec) and high star-to-planet flux ratios (${\sim}10^{9}$ for a Jupiter analog or ${\sim}10^{10}$ for an Earth analog in the visible). Atmospheric turbulence prohibits reaching such high flux ratios on the ground, so observations must be made above the Earth's atmosphere. The Nancy Grace Roman Space Telescope (Roman), set to launch in the mid-2020s, will be the first space-based observatory to demonstrate high-contrast imaging with active wavefront control using its Coronagraph Instrument. The instrument’s main purpose is to mature the various technologies needed for a future flagship mission to image and characterize Earth-like exoplanets. These technologies include two high-actuator-count deformable mirrors, photon-counting detectors, two complementary wavefront sensing and control loops, and two different coronagraph types. In this paper, we describe the complete set of flight coronagraph mask designs and their intended combinations in the Roman Coronagraph Instrument. There are three types of mask configurations included: a primary one designed to meet the instrument's top-level requirement, three that are supported on a best-effort basis, and several unsupported ones contributed by the NASA Exoplanet Exploration Program. The unsupported mask configurations could be commissioned and used if the instrument is approved for operations after its initial technology demonstration phase.
\end{abstract}


\section{INTRODUCTION}
\label{sec:intro}  

Over four thousand exoplanets have been discovered to date, primarily via indirect methods that detect changes in the flux or Doppler shift of the host stars. Several of these worlds are even Earth-sized and lie within the habitable zones of their stars. To determine if any of these or other exoplanets are habitable, we need spectroscopic measurements of their atmospheres. Except for rare cases in which the exoplanet transits its host star, this can only be achieved by directly imaging reflected light from the exoplanet. The direct imaging of exoplanets is challenging because of the high star-to-planet flux ratios at small angular separations. For example, the Sun and Earth viewed from ten parsecs away have a planet-to-star flux ratio of ${\sim}10^{-10}$ in the visible and a maximum separation of just 0.1 arcseconds. Imaging planets with such a high flux ratio will only be possible with a space-based observatory, avoiding the limitations of atmospheric turbulence.

In conjunction with a large space telescope for its collecting area and angular resolution, specialized optics are needed to suppress the bright glare of the starlight at the exoplanet's location. One of the leading technologies for this is the internal coronagraph. A coronagraph is a set of masks and/or mirrors that blocks or redirects the on-axis starlight and transmits off-axis sources such as exoplanets and debris disks. Two flagship mission concepts proposed for the 2020 Astrophysics Decadal Survey, the Habitable Exoplanet Observatory (HabEx)\cite{habex2019final} and the Large Ultraviolet/Optical/Infrared Surveyor (LUVOIR)\cite{luvoir2019final}, would use a coronagraph instrument to image and characterize Earth-like exoplanets around nearby stars.

High-contrast coronagraphs with active wavefront control have been successfully demonstrated in laboratory testbeds but have not yet flown in space. To bridge the the gap between current capabilities and a future Earth-analog characterization mission, NASA is including the Coronagraph Instrument as a technology demonstrator on the Nancy Grace Roman Space Telescope (Roman), a 2.4-meter flagship observatory set to launch in the mid-2020s.\cite{kasdin2020roman, ilya2021roman, mennesson2021roman} Some of the key technologies that the Roman Coronagraph Instrument will demonstrate are: high-order wavefront correction with two high-actuator-count deformable mirrors (DMs),\cite{cady2015spc, seo2016hlc, cady2017demonstration, marx2018spc, seo2017hlc, seo2018hlc} low-order wavefront sensing and control (LOWFS/C) of pointing jitter and thermo-mechanical deformations of the observatory,\cite{shi2016wfirst,shi2017lowfs,shi2019lowfs} photon-counting detectors,\cite{harding2015emccd} and two different types of high-contrast coronagraphs.

The two complementary types of high-contrast coronagraphs included in the Roman Coronagraph Instrument are the Hybrid Lyot Coronagraph (HLC)\cite{moody2007hybrid, trauger2011hybrid, trauger2016hlc, seo2016hlc} and the Shaped Pupil Coronagraph (SPC)\cite{carlotti2011optimal, vanderbei2012fast, zimmerman2016splc, riggs2017splc, riggs2019loworder}. The HLC uses two DMs with large deformations, a phase-and-amplitude modifying focal-plane occulting spot, and a traditional Lyot stop to destructively interfere light and create a high-contrast dark hole in the final image. Much of the remaining starlight diffracts just outside the dark hole, so a field stop is needed with the HLC on Roman to limit the dynamic range on the detector and prevent ghosting through the imaging lens. The SPC uses a binary-amplitude apodizer to reshape the point-spread function (PSF) and then a focal plane mask and Lyot stop to block most of the rest of the light at later planes. Relatively little residual starlight remains at the final image plane, so field stops are not required for the SPCs on Roman in imaging mode, although field stops are required for slit spectroscopy and recommended for polarimetry).

Because of the challenging pupil obscurations (a large central obscuration and thick secondary mirror support struts) of the Roman Space Telescope, multiple coronagraphic mask configurations are needed to enable the desired high-contrast usage cases of 1) imaging close to the star, 2) performing spectroscopy close to the star, 3) imaging over a wide field of view (FOV), and 4) performing polarimetry. The aspirational science cases enabled by these engineering goals are, respectively, to detect cool gas giant exoplanets and image inner debris disks, spectrally characterize the imaged exoplanets, image the outer part of faint debris disks, and measure the polarization of disks.

In this paper, we show all the coronagraphic masks included in the Roman Coronagraph Instrument and all the designed combinations of them. The designed mask configurations are described in Section \ref{sec:mask_configs}. The layouts of the individual masks on their substrates are shown in Appendices \ref{sec:spam}, \ref{sec:fpam}, \ref{sec:lsam}, and \ref{sec:fsam}. The available color filters are listed in Appendix \ref{sec:filters}.


\section{Mask Configurations}
\label{sec:mask_configs}

\subsection{Precision Alignment Mechanisms (PAMs)}
\label{sec:pams}

\begin{figure}[htbp!]
    \centering
    \includegraphics[width=0.5\textwidth,]{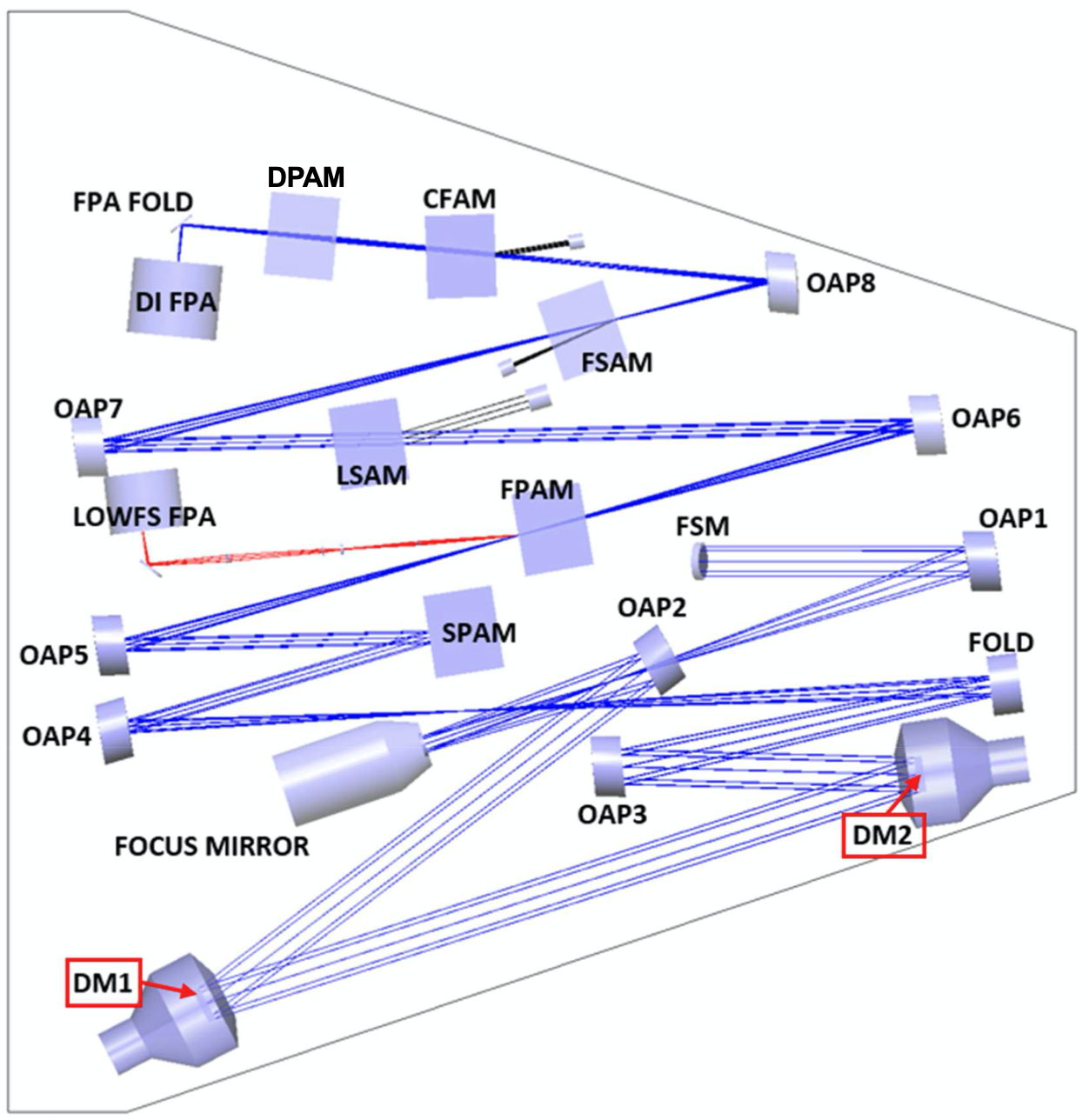}
    \caption[Optical Layout]{The optical layout of the Roman Coronagraph Instrument. The instrument's beam path begins at the fast steering mirror (FSM) and ends at the direct imaging detector (labeled DI FPA).}
    \label{fig:optical_layout}
\end{figure}

As shown in the instrument's optical layout in Fig.~\ref{fig:optical_layout}, the coronagraphic masks reside in four planes on different Precision Alignment Mechanisms (PAMs), which are the Shaped Pupil Alignment Mechanism (SPAM), Focal Plane Alignment Mechanism (FPAM), Lyot Stop Alignment Mechanism (LSAM), and Field Stop Alignment Mechanism (FSAM). For completeness, in Appendices \ref{sec:spam}, \ref{sec:fpam}, \ref{sec:lsam}, and \ref{sec:fsam} we show the mask PAMs and the layouts of all the masks on their substrates. Most of the mask layouts have yet to be fabricated, so the final products may differ slightly (e.g., extra alignment fiducials might be added) in appearance from what is shown here.

The two other PAMs, located downstream of the coronagraph masks, are the Color Filter Alignment Mechanism (CFAM) and the Dispersion Polarizer Alignment Mechanism (DPAM). The CFAM contains four main bandpass filters: Band 1 is a 10.1\% bandwidth\footnote{Bandwidth is defined here as the full width at half maximum (FWHM) of the filter transmission profile.}~centered at 575 nm; Band 2 is a 17.0\% bandwidth centered at 660nm; Band 3 is a 16.7\% bandwidth at 730nm; and Band 4 is a 11.4\% bandwidth at 825nm. Within each full band, there are also several narrower-bandwidth engineering filters used for digging the dark holes. The complete list of CFAM filters and their designed bandpasses is in Table \ref{tab:filters}, and their final, manufactured properties will be updated in the list of instrument parameters at \url{https://roman.ipac.caltech.edu}.

The DPAM contains several standalone lenses as well as two polarization modules (with Wollaston prisms) and two spectroscopy modules (with Amici prisms).\cite{groff2020prism} The polarization modules are planned for use only in Bands 1 and 4 but could be used without performance degradation in Bands 2 and 3. One spectroscopy module is designed for zero optical deviation and a spectral resolution of 50 at the center wavelength of Band 2, and the other is designed to have those properties at the center of Band 3. The spectroscopy modules could be used in Bands 1 and 4 as well but would have nonzero (but minimal) optical deviation and different spectral resolutions.

\subsection{Classes of Mask Configurations}
\label{sec:classes}

Based on their levels of testing and software support, there are three different classes of mask configurations in the Roman Coronagraph Instrument: primary, supported, and unsupported. 

The primary mask configuration will fulfill the instrument's single, top-level threshold requirement and will therefore receive full, system-level testing on the ground. This top-level requirement is to perform broadband, high-contrast imaging in Band 1 at a small angular separation. It states:
\begin{displayquote}
The Roman Coronagraph shall be able to measure, with SNR  $\geq 5$, the brightness of an astrophysical point source located between 6 and 9 $\lambda$/D from an adjacent star with a V$_{\mathrm{AB}}$ magnitude $\leq 5$, with a flux ratio $\geq 1 \times 10^{-7}$. The bandpass shall have a central wavelength $\leq 600$ nm and a bandwidth $\geq 10$\%.
\end{displayquote}
To mitigate the programmatic risks associated with demonstrating a high-contrast coronagraph with active wavefront control in space for the first time, this threshold requirement is intentionally set to be easier than the current best estimates of instrument performance, which predict broadband dark holes with raw contrasts in the $10^{-9}$ to $10^{-8}$ range.\footnote{Performance predictions for the primary mask configuration and two of the supported ones are available at \url{https://github.com/nasavbailey/DI-flux-ratio-plot} and are updated periodically.} Note that Band 1 polarimetry is supported on a best-effort basis because it is not tied directly to the top-level requirement.

The three supported mask configurations serve to perform spectroscopy in Bands 2 and 3 as well as wide FOV imaging and polarimetry in Band 4. Being part of the plan but not tied to requirements, these supported mask configurations are included on a best-effort basis. They are receiving component-level hardware support and full software support. The main difference is that they will not receive end-to-end, system-level testing prior to launch. 

The NASA Exoplanet Exploration Program (ExEP) is contributing another twenty unsupported mask configurations. The masks for these are being placed in otherwise unused space alongside the supported masks. Other than being installed and aligned, they will not receive any hardware testing or software support as part of the planned technology demonstration. The unsupported mask configurations could be commissioned and used if the instrument is approved for operations after its initial technology demonstration phase. 

The unsupported mask configurations are divided into three sub-categories: high-contrast, low-contrast, and calibration. The unsupported high-contrast configurations are largely chosen to fill gaps in the covered parameter space of capabilities; for example, providing narrow FOV imaging to Band 4. The unsupported low-contrast configurations are all simply Lyot coronagraphs that can utilize the instrument's full field of view. Finally, the unsupported calibration masks are Zernike wavefront sensors (ZWFS) to enable sensitive pupil-plane wavefront measurements, as is desired for the future mission concepts HabEx and LUVOIR.\cite{ruane2020zwfs}

\subsection{Designed Mask Configurations}
\label{sec:designed}

The complete set of designed mask configurations in the Roman Coronagraph Instrument is shown in Table \ref{tab:master}. (Other possible mask configurations are discussed in Section \ref{sec:other}.) Each row in Table \ref{tab:master} is for a different mask configuration, and the rows are grouped into the different classes of mask configurations (primary, supported, and the three types of unsupported). The columns indicate key optical elements and FOV parameters for each configuration.


\begin{table}[hp!]
\begin{adjustbox}{max width=\textwidth}
\begin{tabular}{| m{0.8in} | m{0.35in}  | m{1in} | m{0.3in} | m{0.4in} | m{0.7in} | m{0.4in} | m{1in} | m{0.5in} | m{0.35in} | m{0.4in} | m{0.35in} | m{0.4in} | }
\hline
Configuration Class & Coro. Type & Configuration Name & Band No. & SPAM Pos'n  & FPAM Pos'ns  & LSAM Pos'n & FSAM Positions  & Az FOV (deg) & IWA ($\lambda$/D) & IWA (mas) & OWA ($\lambda$/D) & OWA (mas) \\
\hline
Primary    & HLC                             & Narrow-FOV                      & 1           & 1 & 4-C2R1:9, 4-C3R1:8, 4-C4R7:9          & 3         & R1C1, (R1C3), (R6C6), R5C3, R5C4         & 360             & 3.0                           & 150.6             & 9.7    & 486.8  \\
\hline
\multirow{8}{0.8in}{Supported}        & SPC                       & Bowtie                     & 2           & 4 & 2-R3C2, 2-R4C2, 2-R4C4                & 5         & none, R2C5, R3C1, R3C2, R6C5       & 130             & 3.0                           & 172.8     & 9.1  & 524.2     \\
\cline{2-13}
                     & SPC                    & Bowtie                     & 3           & 4 & 6-R2C2, 6-R4C3, 6-R6C1                & 5         & none,   R1C2, R3C1, R3C2, R4C2, R4C6 & 130             & 3.0                           & 191.2          & 9.1                           & 579.8      \\
\cline{2-13}  
                     & SPC                     & Wide-FOV        & 4           & 2 & 6-R5C1, 6-R8C1, 6-R8C3                & 4         & none, R1C5, R6C1, R6C2             & 360             & 5.9                           & 424.9                         & 20.1                          & 1447.4     \\
\hline

\multirow{8}{0.8in}{Unsupported (High Contrast)} & SPC                     & Rotated bowtie                     & 2           & 3 & 2-R3C1,   2-R3C3, 2-R4C3                & 1         & R4C3,   R4C4                         & 130             & 3.0                           & 172.8                         & 9.1                           & 524.2     \\
\cline{2-13}
                     & SPC                     & Rotated bowtie                     & 3           & 3 & 6-R2C3:4,   6-R4C2                       & 1         & R2C2,   R4C2                         & 130             & 3.0                           & 191.2                         & 9.1                           & 579.8    \\
\cline{2-13}
                      & SPC                    & Wide-FOV                         & 1           & 2 & 2-R1C1:2, 2-R2C1                       & 4         & none, R1C5, R6C1, R6C2             & 360             & 5.9                           & 296.1                         & 20.1                          & 1008.8     \\
\cline{2-13}
                     & SPC                     & Multi-star                             & 1           & 5 & 2-R1C1:2, 2-R2C1                       & 4         & R4C1                                 & 360             & 5.9                           & 296.1                         & 20.1                          & 1008.8     \\
\cline{2-13}
                     & SPC                     & Multi-star                             & 4           & 5 & 6-R5C1,  6-R8C1, 6-R8C3                & 4         & R1C4,   (R4C1)                         & 360             & 5.9                           & 424.9                         & 20.1                          & 1447.4    \\
\cline{2-13}
                     & HLC                     & Narrow-FOV                     & 2           & 1 & 4-C4R1:6,   4-C5R1:8                     & 3        & R3C3, R2C3                     & 360             & 3.0                           & 172.8                         & 9.7                           & 558.8     \\
\cline{2-13}
                     & HLC                     & Narrow-FOV                     & 3           & 1 & 3-R5C1:3,   3-R6C1:3, 3-R7C1:2, 3-R8C1 & 3         & R3C4, R4C5, R5C1, R5C2 & 360             & 3.0                           & 191.2                         & 9.7     & 618.1 \\
\cline{2-13}
                     & HLC    & Narrow-FOV                      & 4           & 1 & 3-R2C1:3, 3-R3C1:3, 3-R4C1:3          & 3         & R3C5                                 & 360             & 3.0                           & 216.0                         & 9.1                           & 655.3     \\
\hline
\multirow{10}{0.8in}{Unsupported (Low Contrast)}   & Lyot         & 0.281" occulter                               & 1, 2        & 1 & 4-C2R1:9, 4-C3R1:8, 4-C4R7-9 & 3, 4 & open, any                            & 360             & -          & 140.5   & -    & 3600   \\ \cline{2-13}
                       & Lyot                    & 0.322" occulter                              & 1, 2        & 1 & 4-C4R1:6, 4-C5R1:8                     & 3, 4 & open, any                            & 360             & -                             & 161.1                         & -                             & 3600   \\ \cline{2-13}
                       & Lyot                   & 0.803" occulter                                & 1, 2        & 1 & 4-C1R3:4                                  & 3, 4 & open, any                            & 360             & -                             & 401.5                         & -                             & 3600 \\ \cline{2-13}
                      & Lyot                    & 0.921" occulter                                & 1, 2        & 1 & 4-C1R2                                    & 3, 4 & open, any                            & 360             & -                             & 460.7                         & -                             & 3600  \\ \cline{2-13}
                      & Lyot                    & 1.807" occulter                               & 1, 2        & 1 & 4-C1R1                                    & 3, 4 & open, any                            & 360             & -                             & 903.4                         & -                             & 3600  \\ \cline{2-13}
                        & Lyot                  & 0.357" occulter                              & 2,3,4     & 1 & 3-R5C1:3,   3-R6C1:3, 3-R7C1:2, 3-R8C1 & 3, 4 & open, any                            & 360             & -                             & 178.6                         & -               & 3600   \\ \cline{2-13}
                        & Lyot                  & 0.403" occulter                              & 2,3,4     & 1 & 3-R2C1:3,   3-R3C1:3, 3-R4C1:3          & 3, 4 & open, any                            & 360             & -                             & 201.6          & -        & 3600   \\ \cline{2-13}
                         & Lyot                 & 1.020" occulter                                & 2,3,4     & 1 & 3-R1C3                                    & 3, 4 & open, any                            & 360             & -                             & 509.8                         & -    & 3600     \\ \cline{2-13}
                          & Lyot                & 1.152" occulter                                & 2,3,4     & 1 & 3-R1C1                                    & 3, 4 & open, any                            & 360             & -                             & 576.1                         & -     & 3600   \\ \cline{2-13}
                        & Lyot                  & 2.294" occulter                               & 2,3,4     & 1 & 3-R1C2                                    & 3, 4 & open, any                            & 360             & -                             & 1147.1                        & -  & 3600    \\
\hline
\multirow{2}{0.8in}{Unsupported   (Calibration)}   & ZWFS & transmissive and reflective       & 1           & 1 & 4-C7R2:8                                  & 2         & open                                 & N/A             & N/A                           & N/A                           & N/A                           & N/A      \\ \cline{2-13}
                       & ZWFS & reflective only & N/A           & 1 & 2-R2C2:4, 6-R4C1, 6-R6C2, 6-R8C2 & N/A         & N/A                                 & N/A             & N/A                           & N/A                           & N/A                           & N/A  \\
\hline          
\end{tabular}
\end{adjustbox}
\caption{List of designed mask configurations in the Roman Coronagraph Instrument. The unsupported mask configurations could be commissioned and used only if the instrument is approved for operations after its initial technology demonstration phase. Other combinations of masks are possible but most would have worse contrast, throughput, and/or FOV compared to the configurations listed above. Imaging, spectroscopy, and polarimetry are technically possible in all four bandpasses; however, polarimetry is officially supported only in Bands 1 and 4, and spectroscopy is officially supported only in Bands 2 and 3. Spectroscopy and polarimetry cannot be done concurrently. Mask PAM positions reference the labeled mask array layouts shown in the Appendices. As an example for FPAM,  4-C2R1:9 means column 2 and rows 1 to 9 in overall FPAM position 4. For the SPC Multi-star, the FOV is set by the field stop opening but can be placed anywhere within the listed annular FOV.
}
\label{tab:master}
\end{table}

Column 4 of Table \ref{tab:master} specifies the bandpasses in which each mask configuration can work. Only the full bandpasses are listed, but each mask configuration also works in all the subbands of the full bands listed; the filter bandwidths are given in Table \ref{tab:filters}. All of the high-contrast mask configurations work only in their designed bandpass since they require high-order wavefront correction. The unsupported, low-contrast Lyot coronagraphs do not need wavefront correction and are thus spectrally limited only by the anti-reflective (AR) coatings used on the glass focal plane mask (FPM) substrates. The unsupported Zernike wavefront sensor (ZWFS) mask configurations work in reflection with the low-order wavefront sensor (LOWFS), which has its own 128nm-wide filter centered at 575nm. The ZWFS that also works in transmission is designed to work in Band 1 only. It may work in other bands, but that has not been studied. 

Columns 5-8 of Table \ref{tab:master} specify PAM mask positions that combine to form that mask configuration. There is only one FPM design per mask configuration; the multiple FPAM locations listed per row are all copies of that given FPM design, which provides redundancy against manufacturing defects and/or dust contamination over the course of the mission. The HLC occulters--when combined with flat DMs--can be used as part of the simple Lyot coronagraphs with inner working angles $<0.5$ arcsec. There are only one SPAM mask and one LSAM mask per mode. The low-contrast, traditional Lyot coronagraphs can be used with either Lyot stop that has struts like the telesope pupil (Figs.~\ref{fig:ls_spc_wfov} and \ref{fig:ls_hlc_nfov}), but most users would probably prefer the one in LSAM position 4 because it provides higher throughput and a sharper PSF at the expense of some contrast. 

There are usually several field stops that can work with each mask configuration because the field stops are not used in a diffractive manner to help create the dark hole. Only the field stops designed specifically for each mask configuration are listed in each row of column 8; other desirable pairings are left for the reader to determine based on the field stop specifications provided in Table \ref{tab:fs}. Of the 31 field stops at the FSAM, 27 are unique. The positions of the four that are duplicates are shown in parentheses in Table \ref{tab:master}. All the slit-shaped field stops are intended for use with the spectroscopy modules. The two largest circular field stops (having 1.9-arcsecond and 3.5-arcsecond radii) are primarily intended for use with the Wollaston prisms to eliminate crosstalk between polarization states in the event that the astrophysical source or detectable instrumental scattered light is present at large separations. (The field centers of the two polarization state images are separated by 7.5'' on the detector). 

The FOV is specified by the azimuthal coverage, inner working angle (IWA), and outer working angle (OWA). All configurations have 360-degree azimuthal FOV except for the bowtie SPCs, which each have $2\times65^\circ$ coverage. The IWA and OWA are, respectively, the angular separations at which the off-axis throughput ramps up or ramps down to half the maximum value. With the unsupported, traditional Lyot coronagraphs, the IWA column instead lists the occulting spot radius and the OWA column gives the instrument's full, unvignetted FOV in imaging mode. For the multi-star imaging mask configurations,\cite{bendek2021mswc} the field of view is limited to the $9{\times}9$ $\lambda$/D field stop, but that field stop can be placed anywhere within the annular field of view listed in Table \ref{tab:master}.

\begin{figure}[ht!]
    \centering
    \includegraphics[width=1.0\textwidth,]{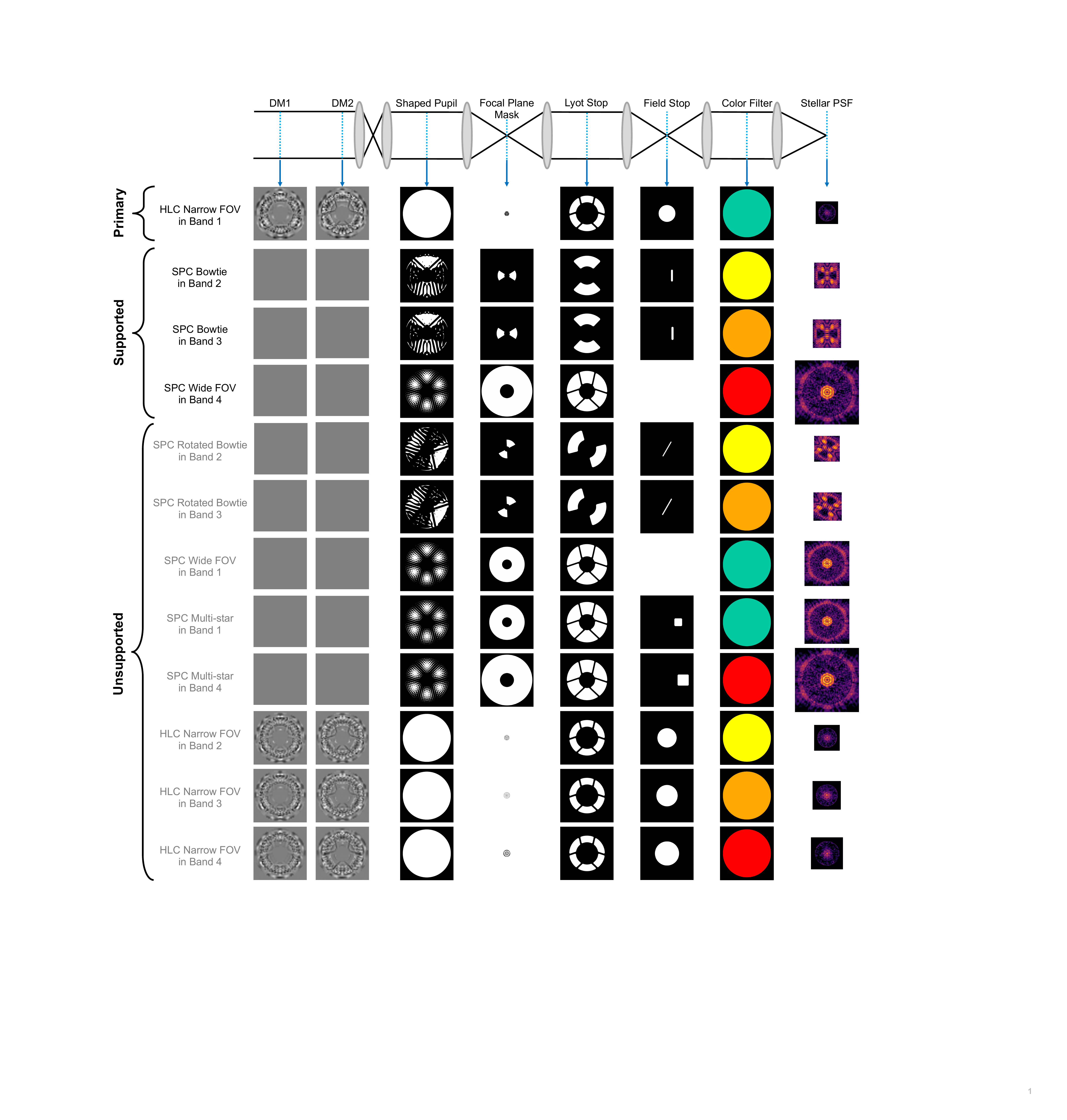}
    \caption[Mask Configuration Grid]{Masks in all the designed high-contrast mask configurations of the Roman Coronagraph Instrument. In general, many field stops could be used with each mask config, but only the nominal field stop for each config is shown above. The DM surfaces shown are in addition to what is needed to correct aberrations in the optical system. The multi-star imaging dark holes are shown for a single star only, as any two-star geometry would be case dependent. All the focal plane masks, field stops, and PSFs are shown to scale relative to each other (in terms of $\lambda/D$). All pupil masks and DM surfaces are shown to scale relative to each other (in terms of beam diameter).}
    \label{fig:mask_grid}
\end{figure}

Figure~\ref{fig:mask_grid} provides a visual representation of the DM surfaces, masks, and stellar PSFs corresponding to all the high-contrast mask configurations listed in Table \ref{tab:master}. All the focal plane masks, field stops, and PSFs are shown to scale relative to each other (in terms of $\lambda/D$). All pupil masks and DM surfaces are also shown to scale relative to each other (in terms of beam diameter).

Looking at the combined capabilities of the primary, supported, and unsupported high-contrast mask configurations, much of the possible parameter space is covered with respect to bandpass, field of view, and spectroscopy and polarimetry. The HLC configurations enable imaging and polarimetry at small separations in all four bandpasses. The SPC wide FOV allows those out to larger separations in Bands 1 and 4. For spectroscopy in Bands 2 and 3, the SPC bowtie and SPC rotated bowtie mask configurations cover approximately two-thirds of the available azimuthal FOV at any given time. The Band 2 and 3 HLCs could also be used for spectroscopy, but the SPC bowties are preferred since they are more robust to the observatory jitter and wavefront drift expected over long integration times.

Another unsupported capability in the instrument is multi-star imaging at high contrast. The apodizer for multi-star imaging (at SPAM position 5) is a near-duplicate of the SPC Wide FOV apodizer (at SPAM position 2) except that it contains a square grid of dots to act as a mild diffraction grating. This creates a grid of bright satellite spots in the focal plane that can be used for super-Nyquist wavefront control (SNWC),\cite{thomas2015snwc} which is a critical component of multi-star wavefront control (MSWC).\cite{sirbu2017mswc} Bendek et al.\cite{bendek2021mswc}~(these proceedings) describe in detail the design and usage cases of the Band 1 and Band 4 multi-star imaging mask configurations in the Roman Coronagraph Instrument.

\subsection{Other Possible Mask Configurations}
\label{sec:other}

Other than the designed mask configurations listed in Section \ref{sec:designed}, there are many other possible combinations of masks and filters. Some may serve a useful purpose while others are highly discouraged due to poor performance. Either way, all of these other possible mask configurations are unsupported.

\subsubsection{Unusable Mask Configurations}

In most instances, arbitrary mask combinations would perform worse in most or all cases compared to the mask configurations already specified.

All the focal plane masks are on glass substrates with AR coatings, which restrict the bandpasses in which they can be used. Inside the intended bandpasses the reflectivity of each substrate surface is $<$0.5\%, but outside it is $>$10\%. Substrates in FPAM positions 2 and 4 can be used in Bands 1 and 2, and substrates in FPAM positions 3 and 6 can be used in Bands 2, 3, and 4.

The SPMs were optimized to make dark holes in their specified bandpasses and fields of view, so decent high-contrast dark holes are not possible out of band or with a different FPM. Because the SPMs increase the brightness of speckles outside the dark hole, one would be better off using the SPAM flat mirror instead of any SPM when performing low-contrast imaging over the instrument's full field of view.

\subsubsection{Other, Viable Mask Configurations}

As mentioned before, the field stops are not diffractive masks (unlike the SPMs, FPMs, and Lyot stops), so there are many field stops that could work for a given combination of other masks. For any high-contrast HLC cases, the dark hole must be dug in a region at least as large as the field stop to prevent bright speckles from reaching the detector (which could damage the detector and would cause ghosting). Because there are 27 unique field stop shapes, we will not describe all the possible field stop usages.

The only known high-contrast mask configurations not already listed would be formed by swapping out one HLC's occulter for another. (The previously mentioned limitation from FPM substrate AR coatings still applies, though.) This is possible because the DMs generate most of the contrast for the HLC, and the DM surfaces are changeable on orbit. A slightly smaller IWA could be obtained (at the expense of worse contrast) in Band 4 by using the Band 3 HLC occulter instead. Similarly, the Band 1 HLC occulter could be used in Band 2. In all other cases, though, using a different HLC occulter (or one of the simple Lyot occulters) would result in a larger IWA (but possibly better contrast). The SPC FPMs could also technically be used with DMs instead of an SPM (similar to the HLCs), but we do not believe these hybrid combinations would work as well as the existing mask configurations.

\section{Summary and Current Status}
\label{sec:summary}

In this paper, we have shown all the coronagraph masks that will be included in the Roman Coronagraph Instrument. We listed and described all the designed mask and color filter combinations. There are four high-contrast mask configurations included as part of the mission's technology demonstration. These configurations accomplish imaging and polarimetry at small angular separations in Band 1, imaging and polarimetry at larger angular separations in Band 4, and imaging and spectroscopy in Bands 2 and 3. The NASA Exoplanet Exploration Program is also contributing a number of unsupported mask configurations that could be used after the initial technology demonstration phase of the mission. These include eight more high-contrast mask configurations to enable multi-star imaging and to complement the FOV and wavelength coverage of the supported mask configurations; several more occulters for low-contrast Lyot coronagraphy; and a Zernike wavefront sensor for high-sensitivity, pupil-plane wavefront estimation.

These masks are all being fabricated by JPL's Microdevices Laboratory.\cite{bala2015fab, bala2017errors, bala2019critical} The masks for the primary mask configuration are being fabricated first and are nearly completed. The remaining masks will be manufactured later in 2021.

\appendix    

\clearpage
\section{Shaped Pupil Masks}
\label{sec:spam}

\begin{figure}[htbp!]
    \centering
    \includegraphics[width=0.8\textwidth,]{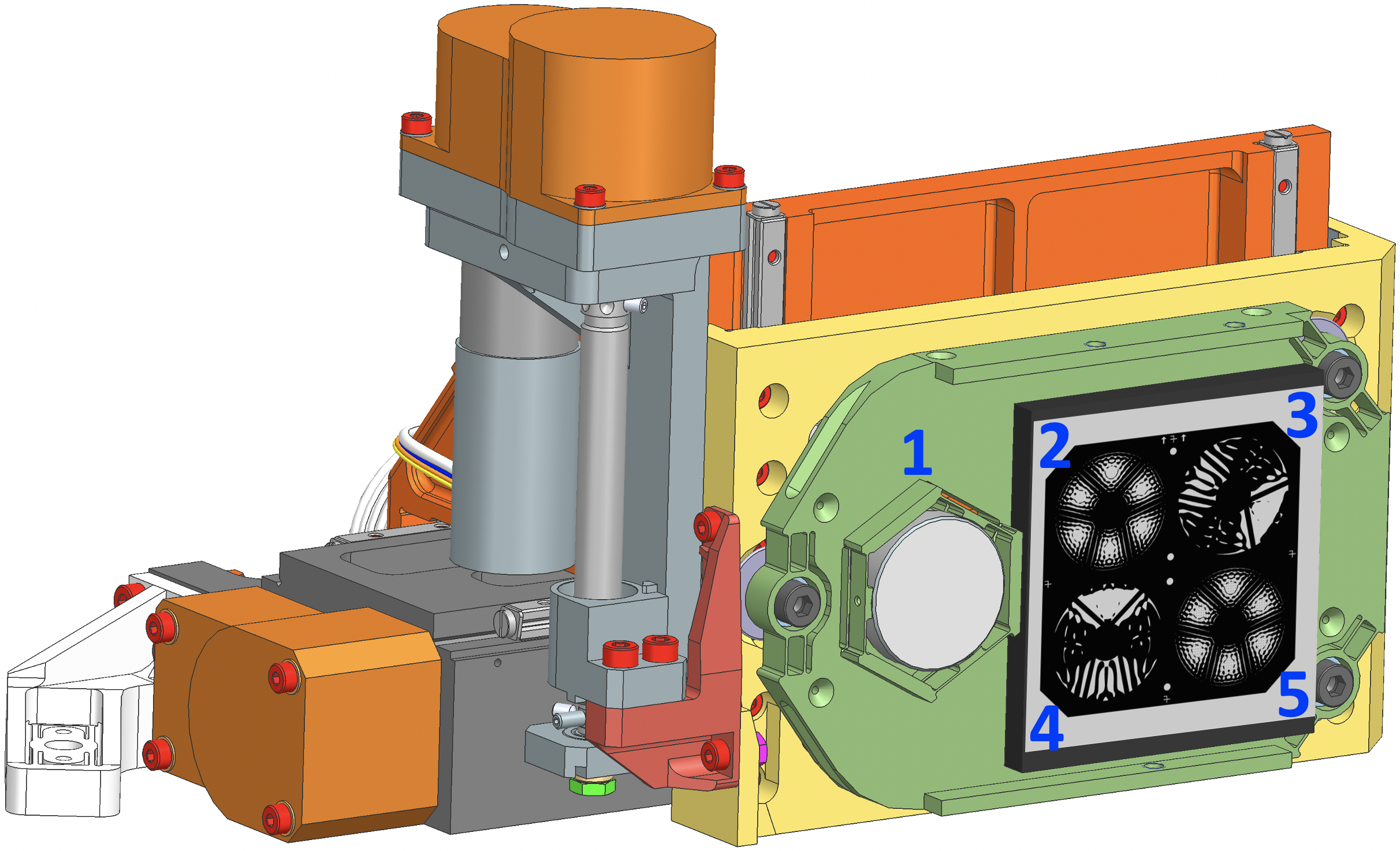}
    \caption[SPAM]{The Shaped Pupil Alignment Mechanism (SPAM) in the Roman Coronagraph Instrument. The five nominal positions are labeled.
    }
    \label{fig:spam}
\end{figure}

\begin{figure}[htbp!]
    \centering
    \includegraphics[width=0.4\textwidth,]{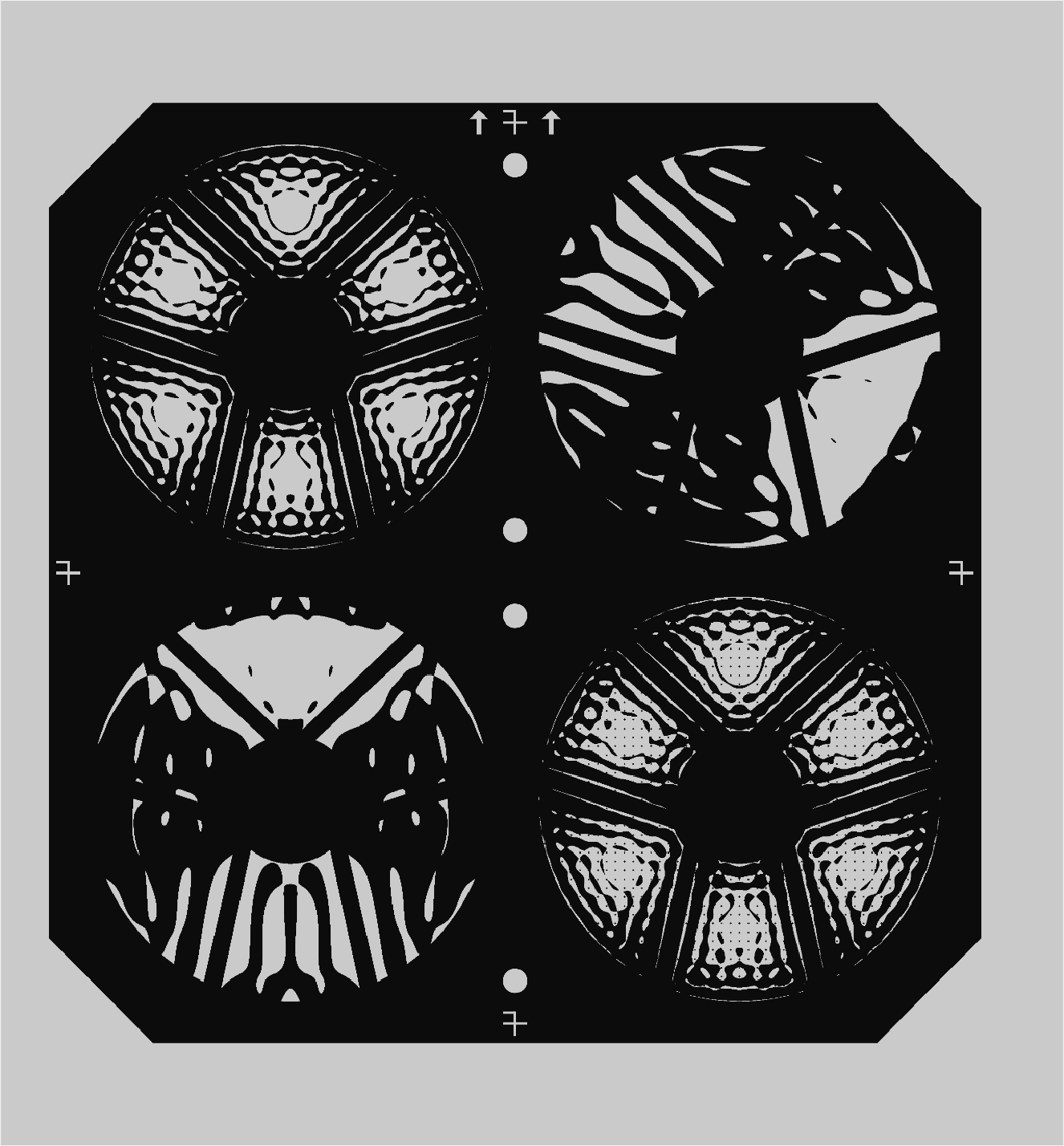}
    \caption[SPM Array]{The array of shaped pupil masks in the Roman Coronagraph. All four masks are manufactured together on a 44mm-wide by 47mm-high by 4mm-thick silicon wafer. The black regions on the substrate front face are black silicon, and the gray areas are unprotected aluminum.\cite{bala2015fab, bala2017errors, bala2019critical} The small circles in between masks are used for calibration and are not illuminated when the beam is centered on the SPMs.}
    \label{fig:spm_array}
\end{figure}

The Shaped Pupil Alignment Mechanism (SPAM), shown in Fig.~\ref{fig:spam}, carries four different shaped pupil masks (SPMs) for the SPCs as well as a standard flat mirror used for the HLCs and calibrations. The HLC flat mirror is in position 1, the SPC wide FOV SPM is in position 2, the SPC rotated bowtie SPM is in position 3, the SPC bowtie SPM is in position 4, and the SPC multi-star imaging SPM is in position 5. A face-on view of the SPM array is provided in Fig.~\ref{fig:spm_array}.

\clearpage
\section{Focal Plane Masks}
\label{sec:fpam}

\begin{figure}[htbp!]
    \centering
    \includegraphics[width=0.95\textwidth,]{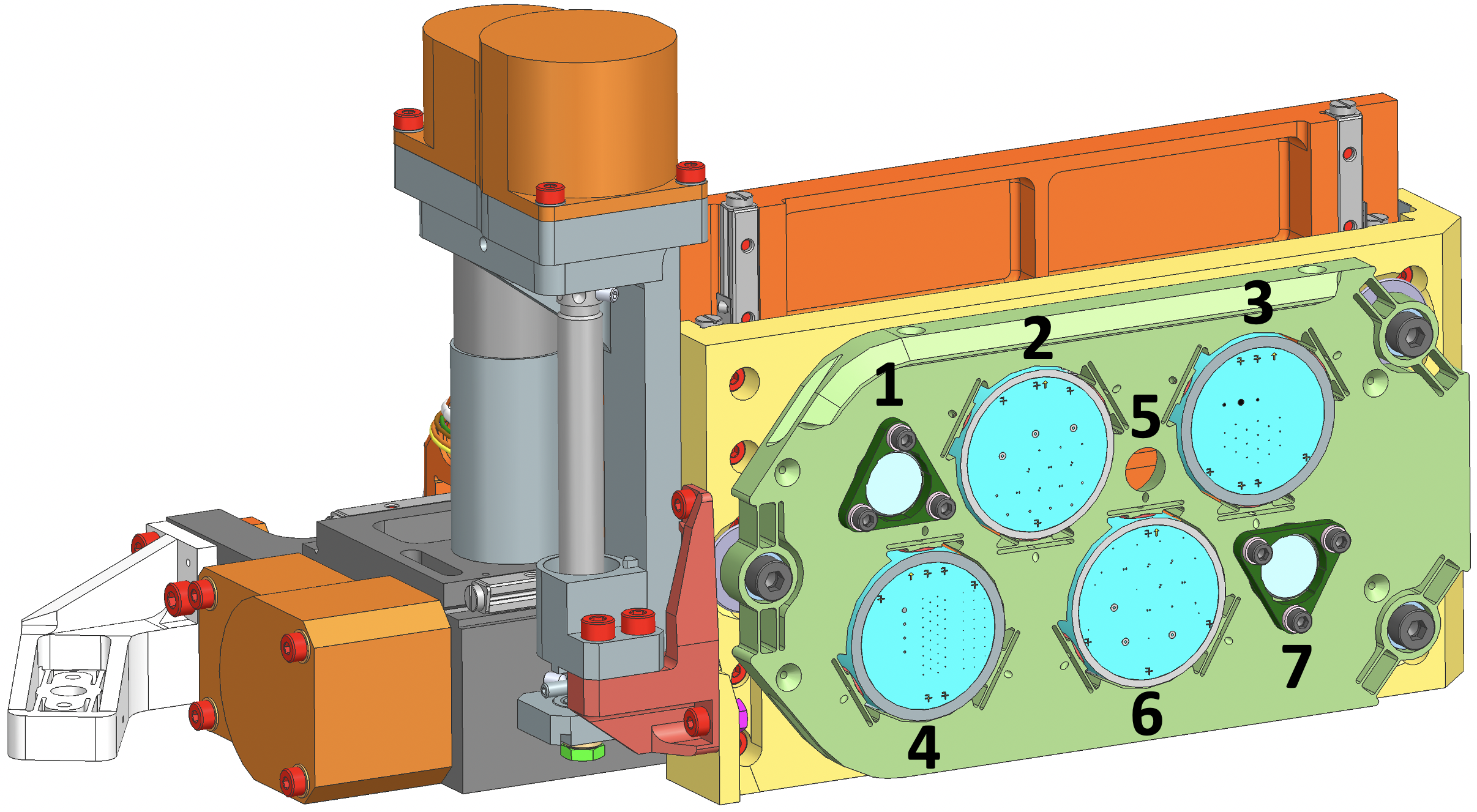}
    \caption[FPAM]{The Focal Plane mask Alignment Mechanism (FPAM) in the Roman Coronagraph Instrument. The mechanism plate has four 23mm-diameter, fused silica substrates carrying the focal plane masks, located at positions 2, 3, 4, and 6. Positions 1 and 7 have neutral-density (ND) filters, and position 5 has a through-hole. 
    }
    \label{fig:fpam}
\end{figure}

\begin{figure}[htbp!]
    \centering
    \includegraphics[width=0.8\textwidth,]{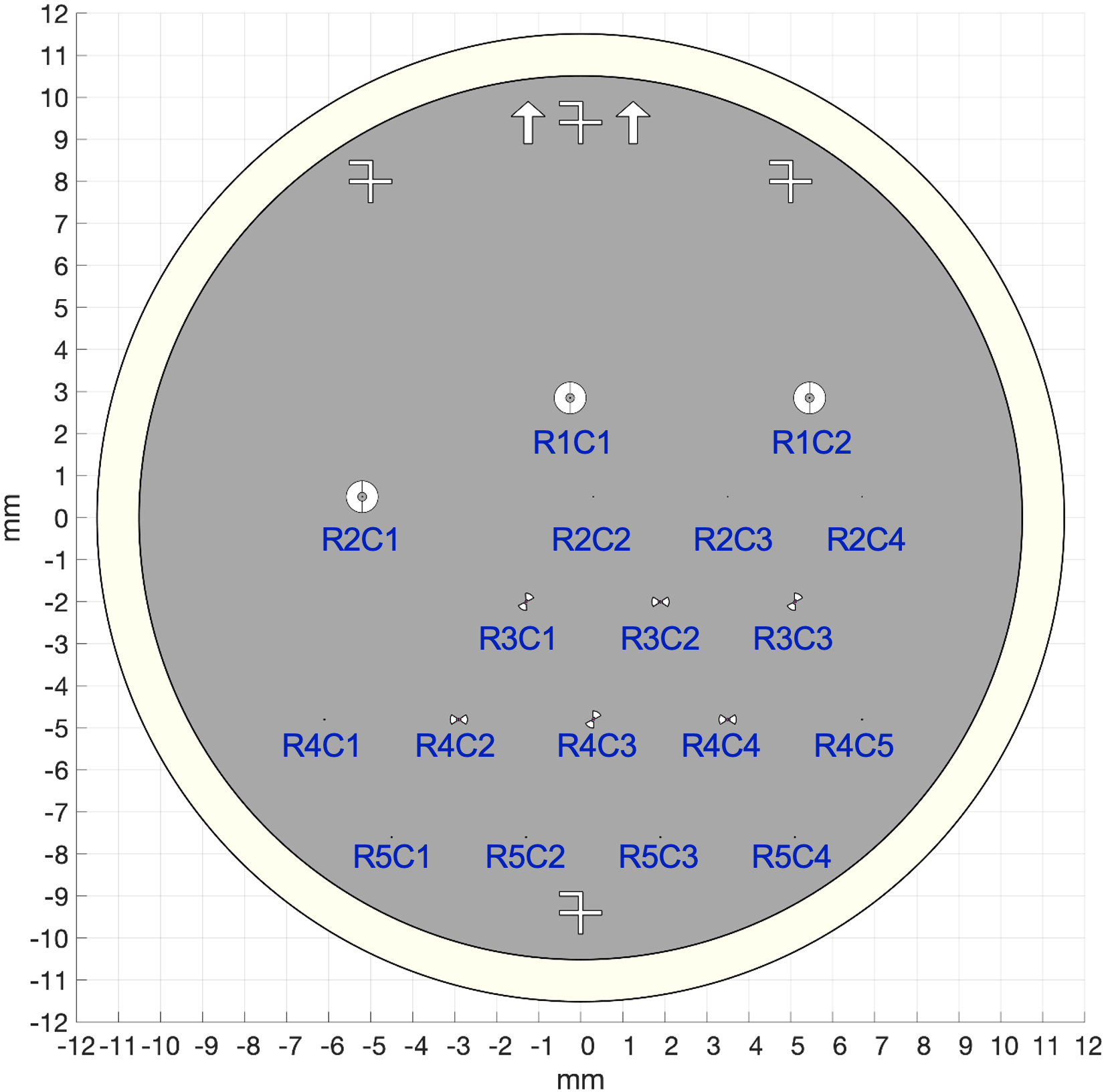}
    \caption[FPAM Position 2]{Mask layout at FPAM position 2, primarily for SPC FPMs in Bands 1 and 2. Note that the vertical line in the annular FPMs is a plotting artifact and is not present in the actual masks.}
    \label{fig:2n}
\end{figure}

\begin{figure}[htbp!]
    \centering
    \includegraphics[width=0.8\textwidth,]{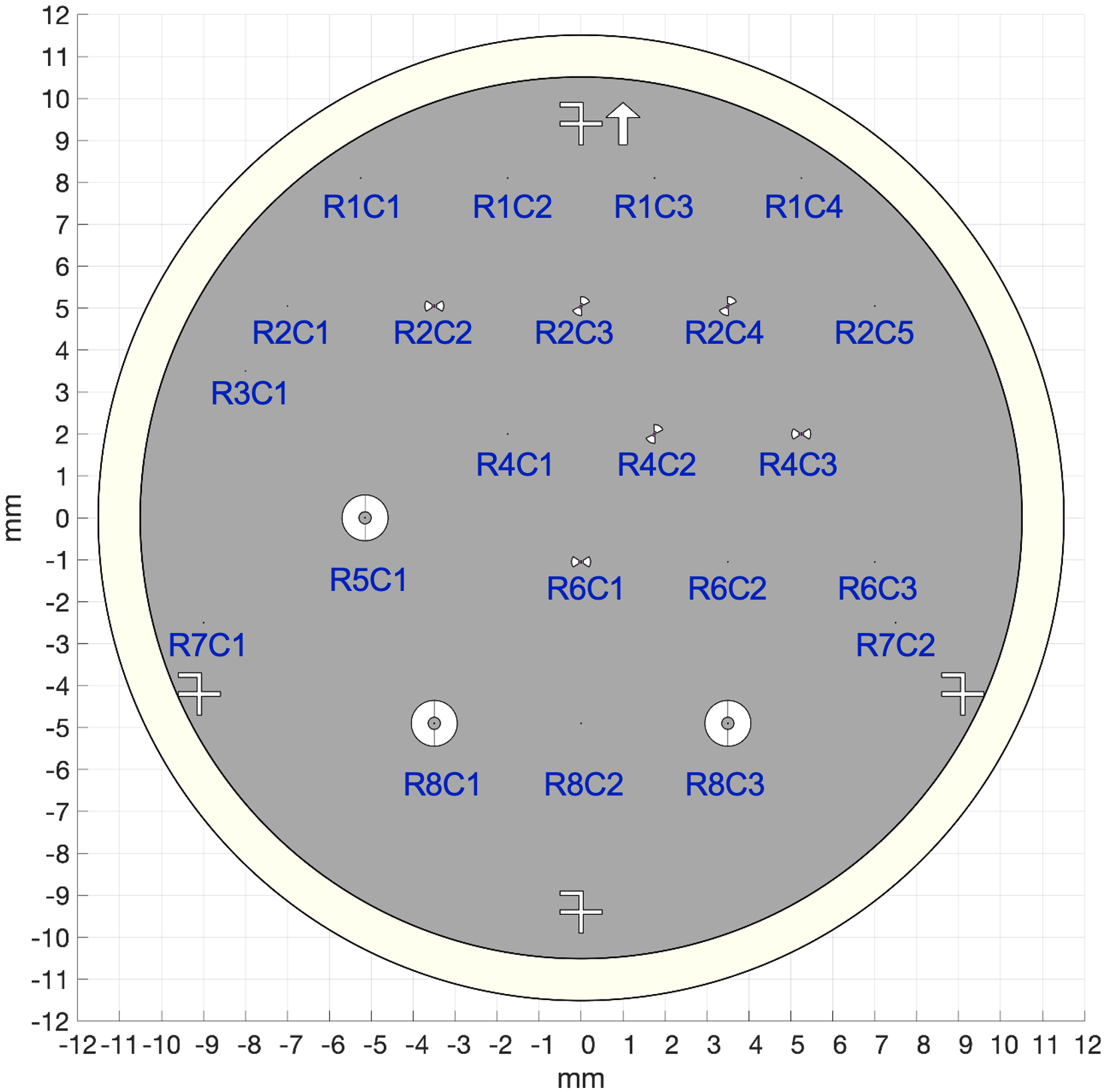}
    \caption[FPAM Position 6]{Mask Layout at FPAM position 6, primarily for SPC FPMs in Bands 3 and 4. Note that the vertical line in the annular FPMs is a plotting artifact and is not present in the actual masks.}
    \label{fig:6n}
\end{figure}

\begin{figure}[htbp!]
    \centering
    \includegraphics[width=0.8\textwidth,]{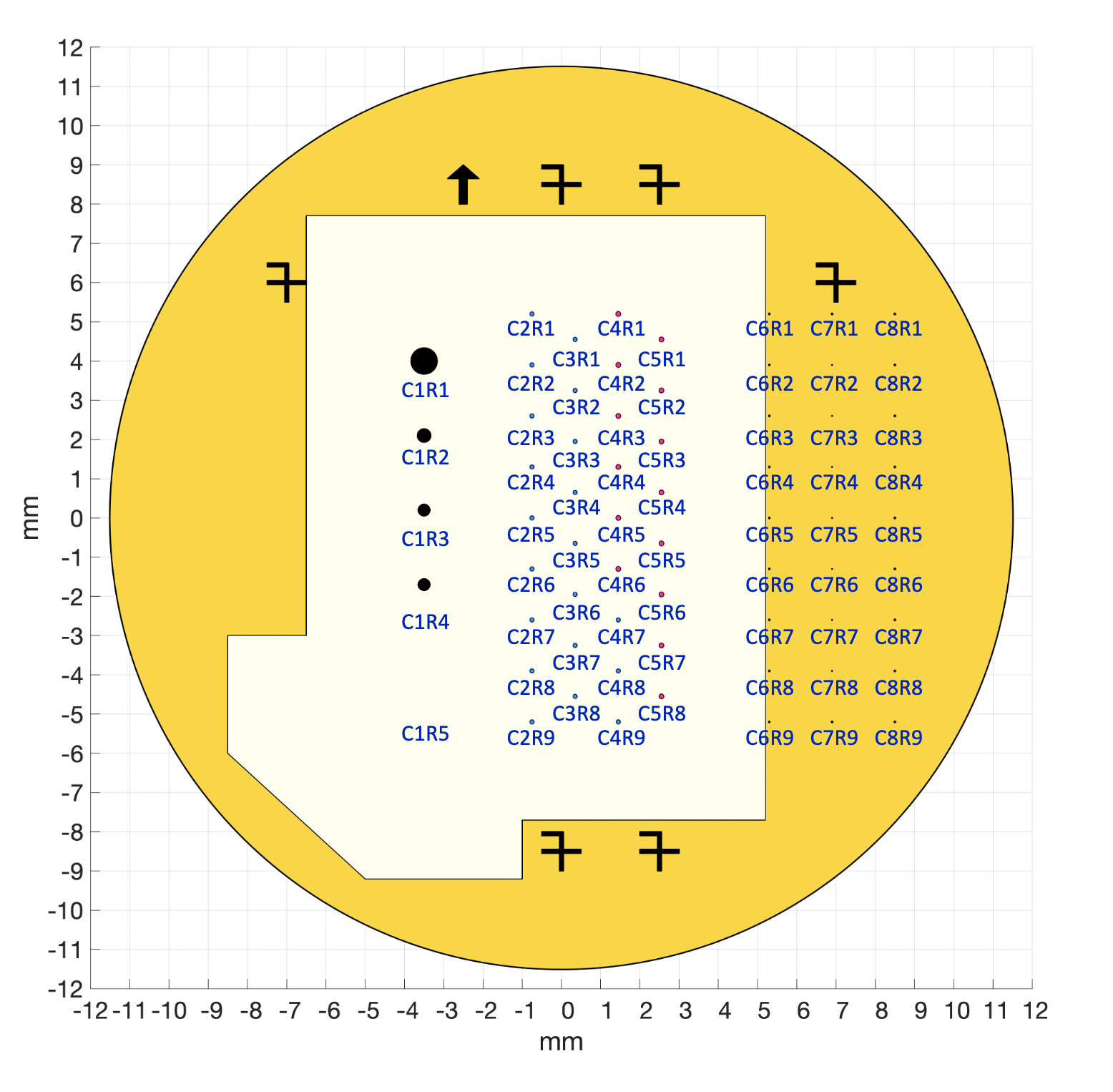}
    \caption[Mask Layout at FPAM Location 4]{Mask Layout at FPAM position 4, primarily for HLC FPMs in Bands 1 and 2. The substrate also includes four simple Lyot coronagraph occulters in column 1 and seven phase-shifting dimples for a Zernike wavefront sensor in column 7. The dark yellow region indicates where PMGI resist is left on the substrate; in general it is not desirable as it negates the AR coating, but here is used to make the ZWFS dimples.}
    \label{fig:4n}
\end{figure}

\begin{figure}[htbp!]
    \centering
    \includegraphics[width=0.8\textwidth,]{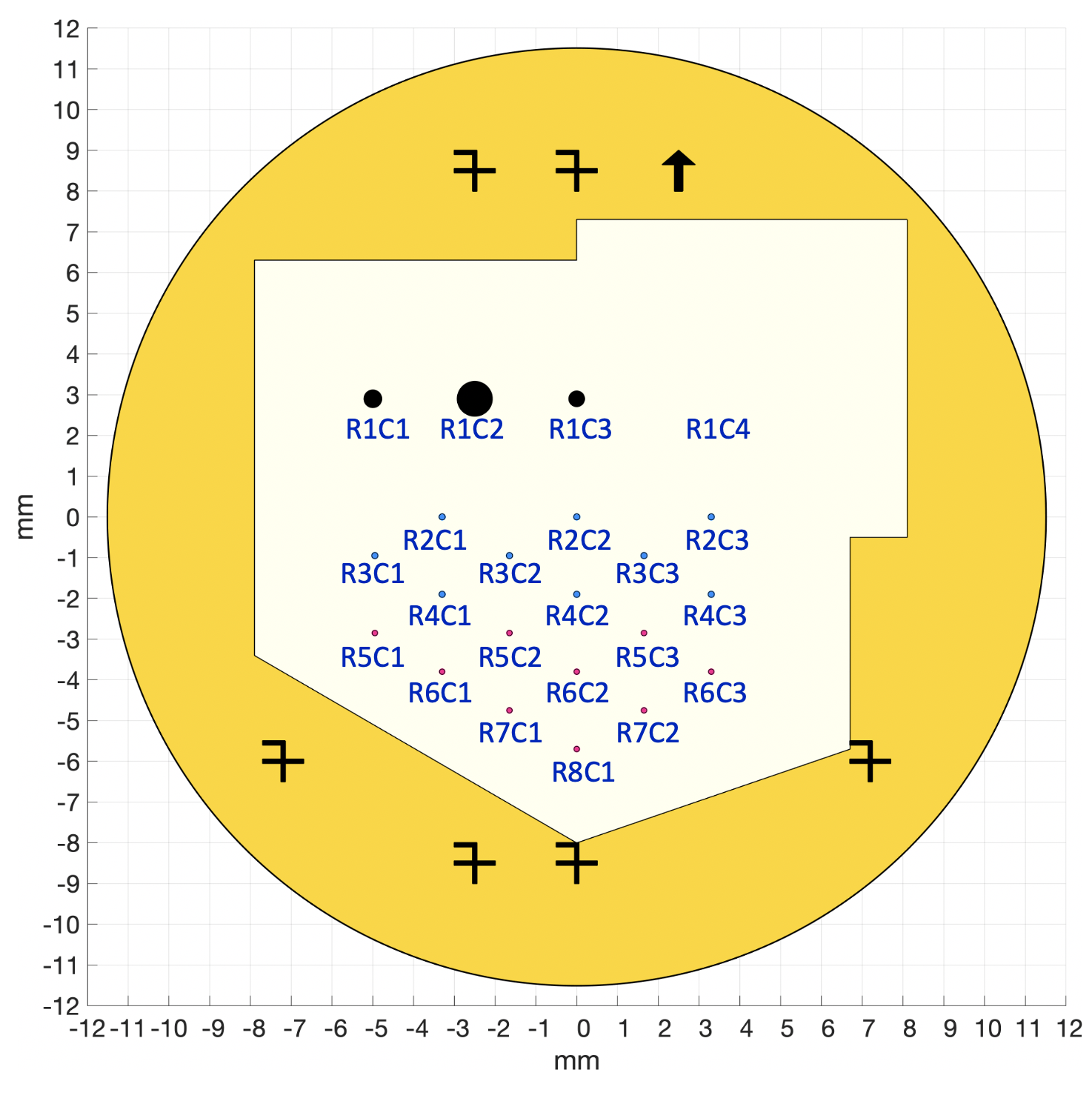}
    \caption[Mask Layout at FPAM Location 3]{Mask layout at FPAM position 3, primarily for HLC FPMs in Bands 3 and 4. The substrate also includes three simple Lyot coronagraph occulters in row 1. All these masks are unsupported. The dark yellow region indicates where PMGI resist is left on the substrate.}
    \label{fig:3n}
\end{figure}

The Focal Plane Alignment Mechanism (FPAM), shown in Fig.~\ref{fig:fpam}, has the most masks of any PAM. FPAM positions 2, 3, 4, and 6 each contain a 23mm-diameter, fused silica substrate carrying many FPMs. Position 2 has FPMs for SPCs in Bands 1 and 2, as shown in Fig.~\ref{fig:2n} and listed in Table \ref{tab:2n}. Position 6 has FPMs for SPCs in Bands 3 and 4, as shown in Fig.~\ref{fig:6n} and listed in Table \ref{tab:6n}. Position 4 has FPMs for HLCs in Bands 1 and 2 as well as ZWFS dimples, as shown in Fig.~\ref{fig:4n} and listed in Table \ref{tab:4n}. Position 3 has FPMs for HLCs in Bands 3 and 4, as shown in Fig.~\ref{fig:3n} and listed in Table \ref{tab:3n}. Several copies of each FPM design are included in case of fabrication defects or dust contamination. The IWA and OWA for each FPM are given in Table \ref{tab:master}.

\clearpage
\begin{table}[htbp!]
\centering
\begin{tabular}{| c | c | c | }
\hline
Mask Name & Band No. & Array Locations \\
\hline
SPC Wide-FOV FPM & 1 & R1C1, R1C2, R2C1 \\
\hline
SPC Bowtie & 2 & R3C2, R4C2, R4C4 \\
\hline
SPC Rotated Bowtie & 2 & R3C1, R3C3, R4C3 \\
\hline
\diameter=0.5$\lambda_1$/D Pinhole & 1 &  R4C1, R4C5, R5C1, R5C2, R5C3, R5C4 \\
\hline
Reflective-only ZWFS Pimple & (LOWFS) & R2C2, R2C3, R2C4 \\
\hline
\end{tabular}
\caption{Mask names and locations on the substrate at FPAM position 2. There are several copies of each mask design. Pinholes are used only during ground calibration. The LOWFS has its own, fixed filter in a separate beam path.}
\label{tab:2n}
\end{table}

\clearpage
\begin{table}[htbp!]
\centering
\begin{tabular}{| c | c | c | }
\hline
Mask Name & Band No. & Array Locations \\
\hline
SPC Wide-FOV FPM & 4 & R5C1, R8C1, R8C3 \\
\hline
SPC Bowtie & 3 & R2C2, R4C3, R6C1 \\
\hline
SPC Rotated Bowtie & 3 & R2C3, R2C4, R4C2 \\
\hline
\diameter=0.5$\lambda_1$/D Pinhole & 1 &  R2C1, R3C1, R7C1 \\
\hline
\diameter=0.5$\lambda_3$/D Pinhole & 3 & R1C4, R2C5, R6C3, R7C2  \\
\hline
\diameter=0.5$\lambda_4$/D Pinhole & 4 & R1C1, R1C2, R1C3  \\
\hline
Reflective-only ZWFS Pimple & (LOWFS) & R4C1, R6C2, R8C2 \\
\hline
\end{tabular}
\caption{Mask names and locations on the substrate at FPAM position 6. There are several copies of each mask design. Pinholes are used only during ground calibration. (Band 1 pinholes are included on the position 6 substrate as a backup because it is supported whereas the position 2 substrate is unsupported.)}
\label{tab:6n}
\end{table}

\clearpage
\begin{table}[htbp!]
\centering
\begin{tabular}{| c | c | c | }
\hline
Mask Name & Band No. & Array Locations \\
\hline
\diameter=1.807 arcsec Lyot Occulter & 1, 2 & C1R1 \\
\hline
\diameter=0.921 arcsec Lyot Occulter & 1, 2 & C1R2 \\
\hline
\diameter=0.803 arcsec Lyot Occulter & 1, 2 & C1R3:4 \\
\hline
Open & 1, 2 & C1R5 \\
\hline
HLC Occulter & 1 & C2R1:9, C3R1:8, C4R7:9 \\
\hline
HLC Occulter & 2 & C4R1:6, C5R1:8  \\
\hline
Transmissive-and-reflective ZWFS Dimple & 1 & C7R2:8 \\
\hline
\diameter=2$\lambda_1$/D Alignment Spots for ZWFS & 1 & C6R1:9, C7R1, C7R9, C8R1:9 \\ 
\hline
\end{tabular}
\caption{Mask names and locations on the substrate at FPAM position 4.}
\label{tab:4n}
\end{table}

\clearpage
\begin{table}[htbp!]
\centering
\begin{tabular}{| c | c | c | }
\hline
Mask Name & Band No. & Array Locations \\
\hline
\diameter=1.152 arcsec Lyot Occulter & 2, 3, 4 & R1C1 \\
\hline
\diameter=2.294 arcsec Lyot Occulter & 2, 3, 4 & R1C2 \\
\hline
\diameter=1.020 arcsec Lyot Occulter & 2, 3, 4 & R1C3 \\
\hline
Open & 2, 3, 4 & R1C4 \\
\hline
HLC Occulter & 4 & R2C1:3, R3C1:3, R4C1:3 \\
\hline
HLC Occulter & 3 & R5C1:3, R6C1:3, R7C1:2, R8C1 \\
\hline
\end{tabular}
\caption{Mask names and locations on the substrate at FPAM position 3.}
\label{tab:3n}
\end{table}

\clearpage
\section{Lyot Stop Masks}
\label{sec:lsam}

\begin{figure}[htbp!]
    \centering
    \includegraphics[width=0.95\textwidth,]{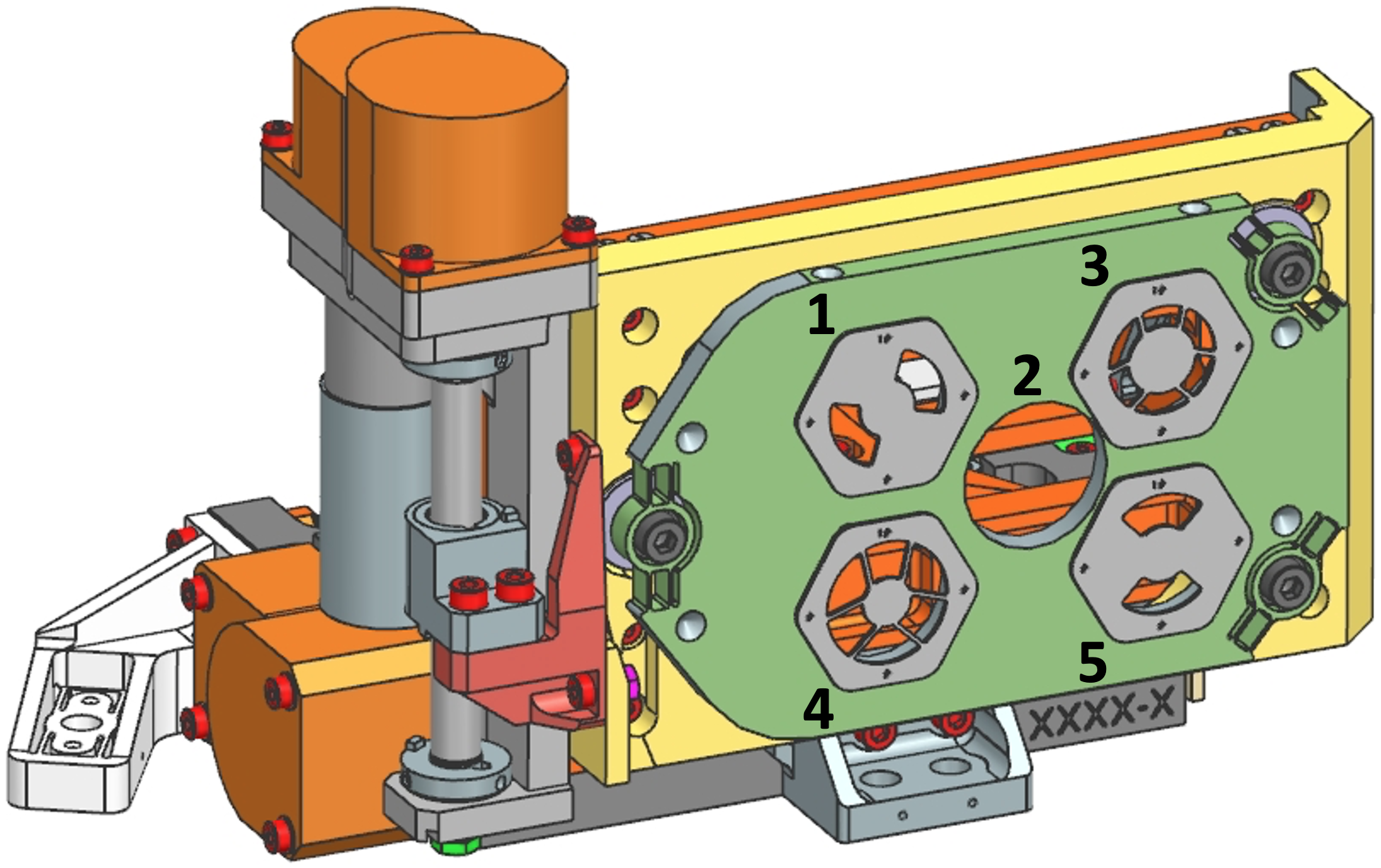}
    \caption[LSAM]{The Lyot Stop Alignment Mechanism (LSAM) in the Roman Coronagraph Instrument. The five nominal positions are numbered.}
    \label{fig:lsam}
\end{figure}

\begin{figure}[htbp!]
    \centering
    \begin{subfigure}[b]{.45\textwidth}
        \centering
        \includegraphics[height=2.0in,]{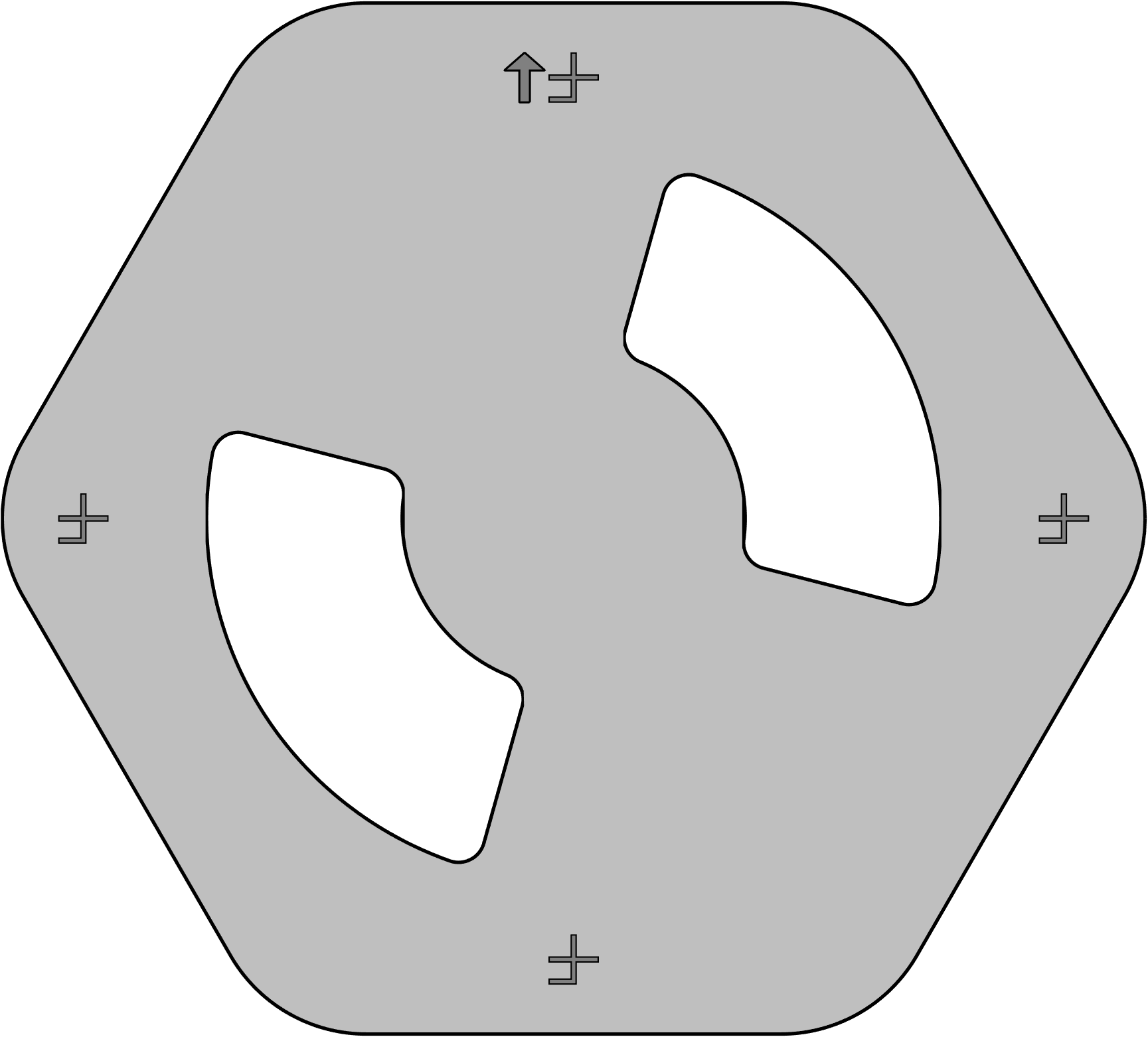}
        \caption{Lyot stop for SPC Rotated Bowtie}
        \label{fig:ls_rotated_bowtie}
    \end{subfigure}
    \begin{subfigure}[b]{.45\textwidth}
        \centering
        \includegraphics[height=2.0in,]{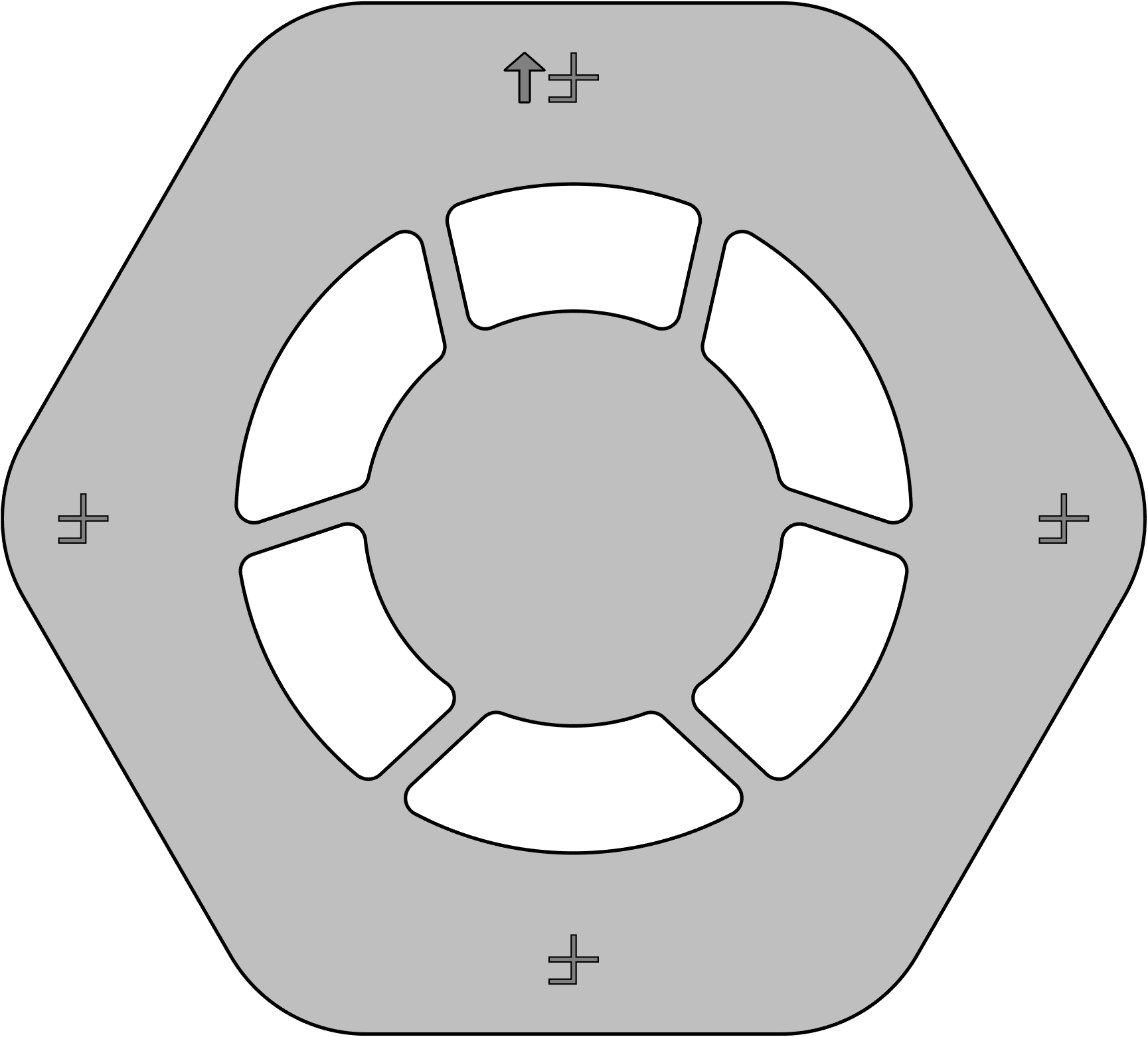}
        \caption{Lyot stop for HLC Narrow-FOV}
        \label{fig:ls_hlc_nfov}
    \end{subfigure}
    \begin{subfigure}[b]{.45\textwidth}
        \centering
        \includegraphics[height=2.0in,]{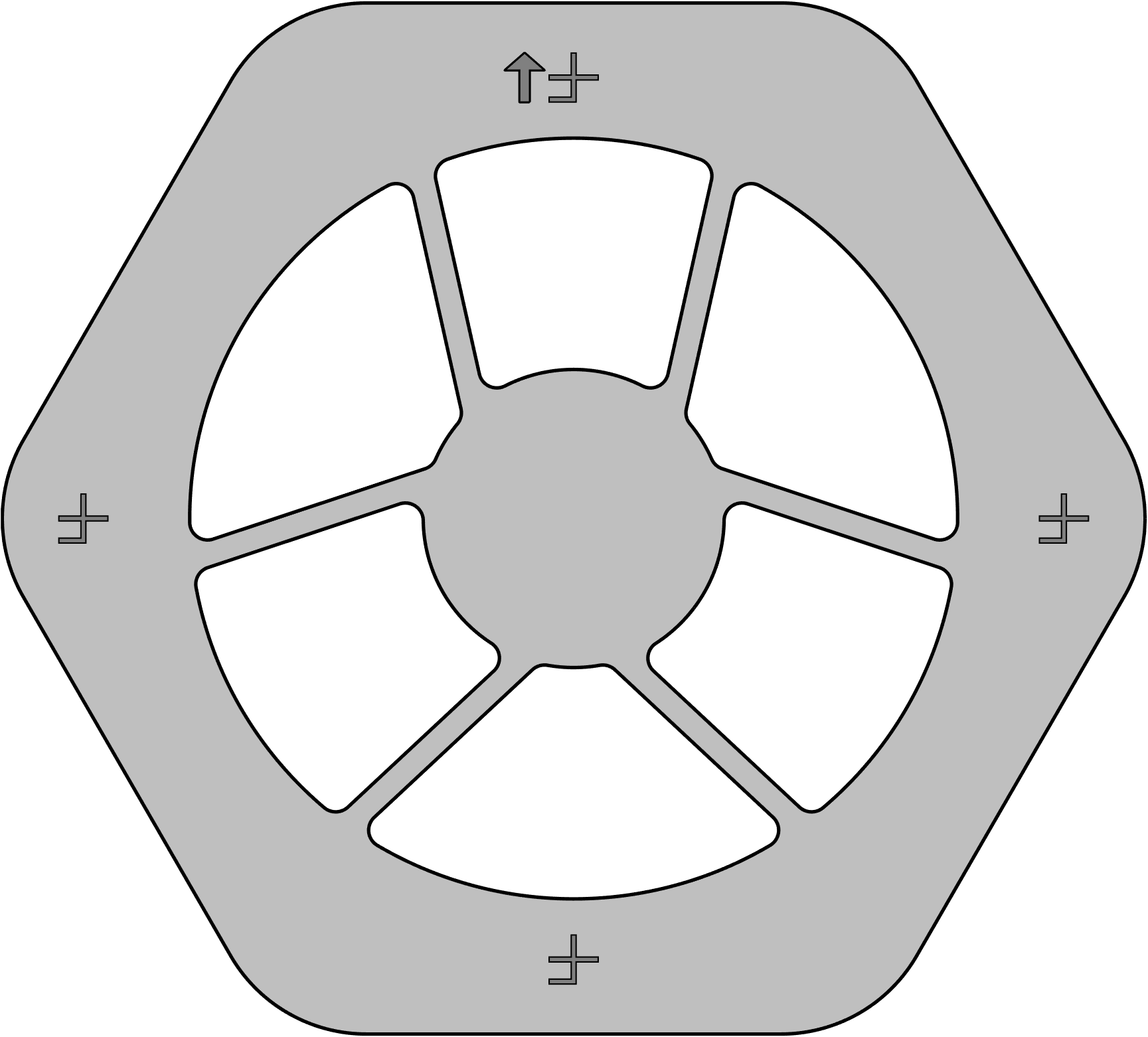}
        \caption{Lyot stop for SPC Wide-FOV and SPC Multi-star}
        \label{fig:ls_spc_wfov}
    \end{subfigure}
    \begin{subfigure}[b]{.45\textwidth}
        \centering
        \includegraphics[height=2.0in,]{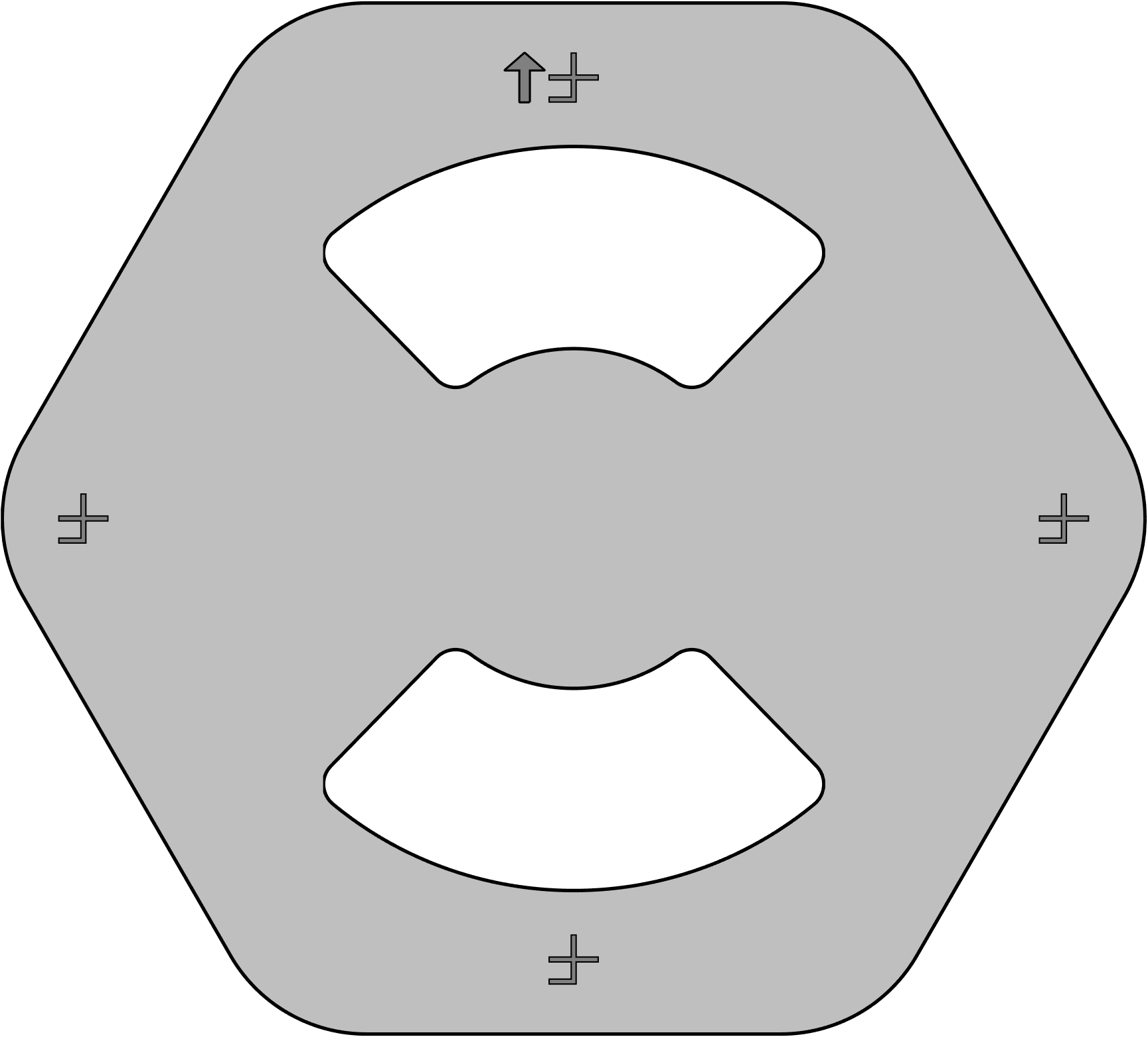}
        \caption{Lyot stop for SPC Bowtie}
        \label{fig:ls_bowtie}
    \end{subfigure}
    \caption[All four Lyot stop masks]{Face-on views of the four Lyot stop substrates. All features are to scale. The flat-to-flat width of each piece is 21.00mm. The fiducials (arrows and rotated F's) are top-layer etches only, not through-holes.}
    \label{fig:lyot_stops}
\end{figure}

The Lyot Stop Alignment Mechanism (LSAM), shown in Fig.~\ref{fig:lsam}, carries four different lyot stops as well as a though hole used for alignment and calibration. The Lyot stop for the unsupported SPC rotated bowtie is in position 1. The through hole is in the middle at position 2. The Lyot stop for all HLCs is in position 3. The Lyot stop for SPC wide FOV and SPC multi-star imaging is in position 4. The Lyot stop for the supported SPC bowtie is in position 5. Face-on views of the Lyot stops are provided in Fig.~\ref{fig:lyot_stops}.

\clearpage
\section{Field Stop Masks}
\label{sec:fsam}

\begin{figure}[htbp!]
    \centering
    \includegraphics[width=0.95\textwidth,]{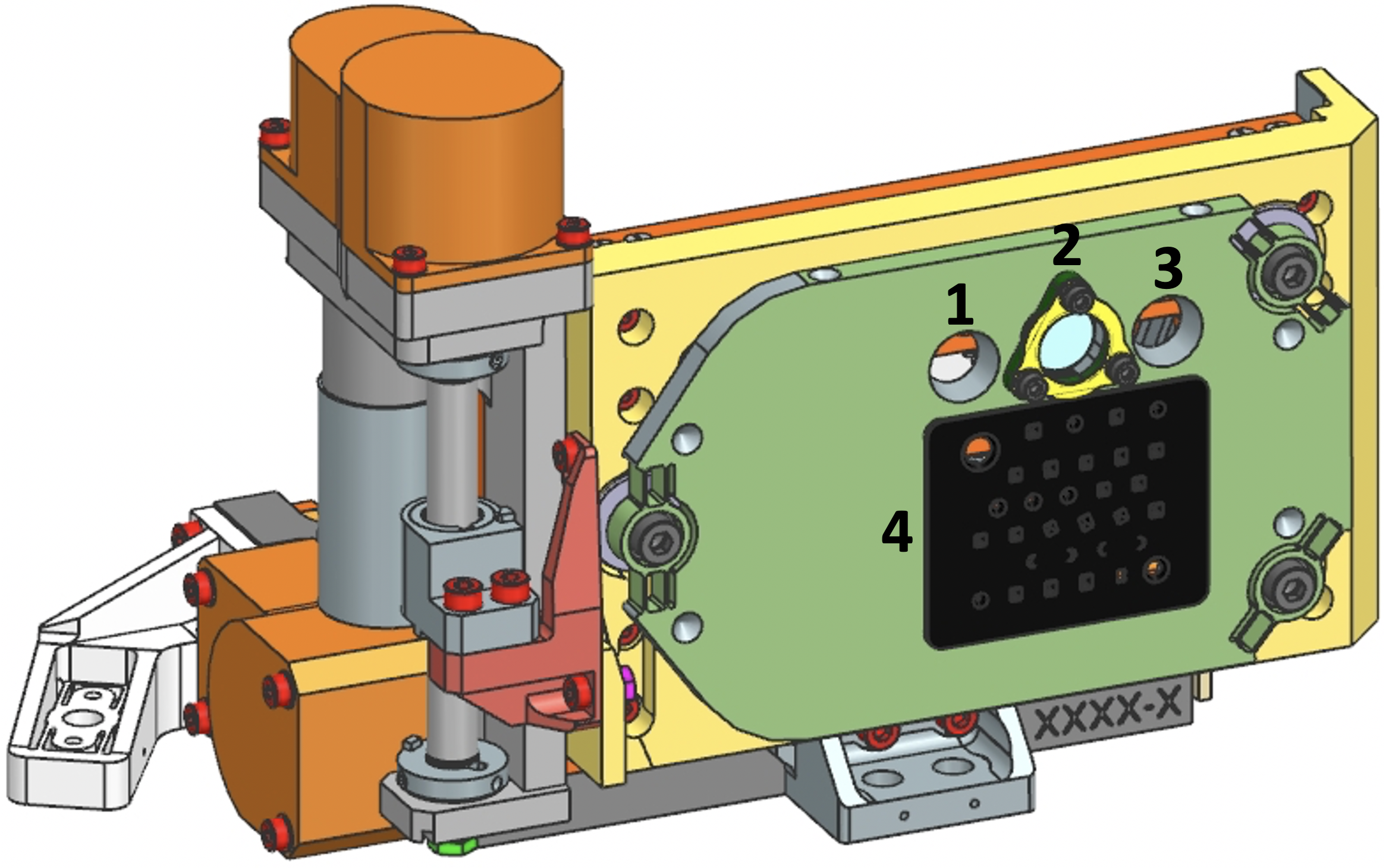}
    \caption[FSAM]{The Field Stop Alignment Mechanism (FSAM) in the Roman Coronagraph Instrument. Unlike the other PAMs, the FSAM is installed facing downstream for packing reasons on the optical bench; thus, the field stop array is mounted face down.}
    \label{fig:fsam}
\end{figure}

\begin{figure}[htbp!]
    \centering
    \includegraphics[width=0.95\textwidth,]{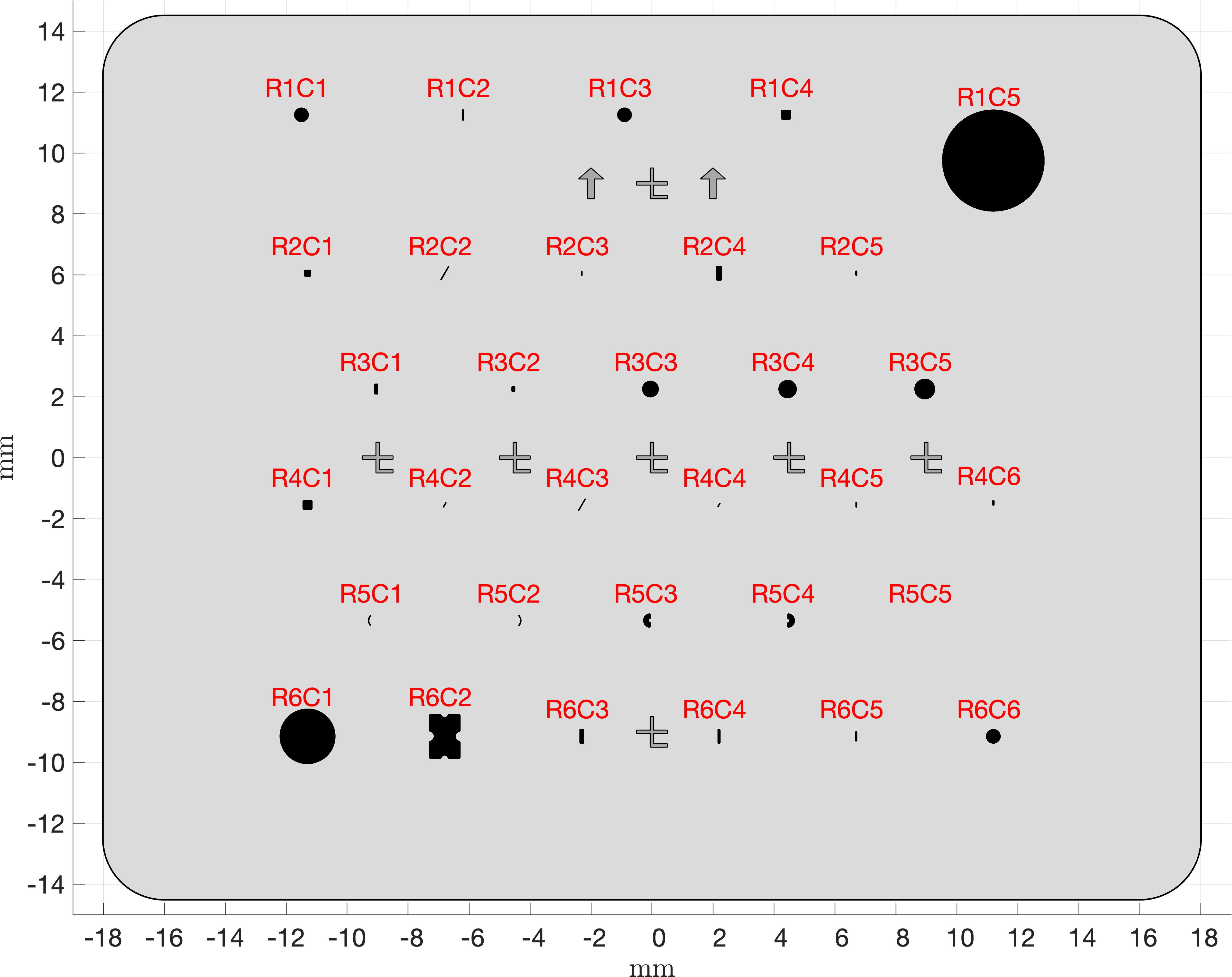}
    \caption[Field stop array]{The field stop array layout at FSAM position 4 in the Roman Coronagraph Instrument. All masks are to scale and are labeled with their position identifiers. The field stops are through holes, whereas the fiducials (arrows and lowercase ``t" shapes) are top-layer etches only. The specifications of each field stop are in Table \ref{tab:fs}.}
    \label{fig:fs_array}
\end{figure}

The Field Stop Alignment Mechanism (FSAM), shown in Fig.~\ref{fig:fsam}, has two open positions (1 and 3), a neutral density filter in position 2, and a silicon wafer carrying all the through-hole field stops in position 4. A face-on view of the substrate with all the field stops is provided in Fig.~\ref{fig:fs_array}. The specifications for all the field stops are listed in Table \ref{tab:fs}.

\begin{table}[htbp!]
\begin{adjustbox}{max width=\textwidth}
\begin{tabular}{ c c c m{0.6in} m{0.5in} m{0.5in} m{0.6in} m{0.5in} m{0.5in} m{0.6in} } 
\hline
FSAM Position & Primary Use  & Shape  & Width ($\mu$m) & Width ($\lambda$/D) & Width (mas) & Height ($\mu$m) & Height ($\lambda$/D) & Height (mas) & Fillet Radii (micron) \\
\hline
1     & Open       & -       & -         & -          & -         & -          & -           & -          & -                   \\
2     & Neutral density (ND) filter       & -       & -         & -          & -         & -          & -           & -          & -                   \\
3     & Open       & -       & -         & -          & -         & -          & -           & -          & -                   \\
4-R1C1     & HLC Band 1        & circle       & 464.34         & 19.4          & 974         & 464.34          & 19.4           & 974          & N/A                   \\
4-R1C2     & SPC major-FWHM vertical slit Band 3; tall       & rect        & 60.77          & 2             & 127         & 364.84          & 12             & 765          & 30.39                 \\
4-R1C3     & HLC Band 1 [spare] & circle         & 464.34         & 19.4          & 974         & 464.34          & 19.4           & 974          & N/A                   \\
4-R1C4     & Multi-star imaging in Band 4                    & square         & 309.07         & 9             & 648         & 309.07          & 9              & 648          & 30                    \\
4-R1C5     & 3.5" radius circular polarimetry field stop     & circle     & 3338.28        & N/A           & 7000        & 3338.28         & N/A            & 7000         & N/A                   \\
4-R2C1     & Multi-star imaging in Band 1                    & square         & 215.41         & 9             & 452         & 215.41          & 9              & 452          & 30                    \\
4-R2C2     & SPC minor-FWHM rotated slit Band 3; tall        & rect           & 30.39          & 1             & 64          & 516.58          & 17             & 1083         & 15.19                 \\
4-R2C3     & HLC FWHM vertical slit Band 2                   & rect           & 30.22          & 1.1           & 63          & 164.84          & 6              & 346          & 15.11                 \\
4-R2C4     & Long slit \#3                              & rect        & 171.68         & N/A           & 360         & 476.9           & N/A            & 1000         & 30                    \\
4-R2C5     & SPC major-FWHM vertical slit Band 2; short      & rect          & 54.95          & 2             & 115         & 164.84          & 6              & 346          & 27.47                 \\
4-R3C1     & SPC $\sim$1st null vertical slit; tall          & rect       & 115.72         & 4             & 243         & 347.16          & 12             & 728          & 30                    \\
4-R3C2     & SPC $\sim$1st null vertical slit; short         & rect       & 115.72         & 4             & 243         & 173.58          & 6              & 364          & 30                    \\
4-R3C3     & HLC Band 2       & circle         & 532.98         & 19.4          & 1118        & 532.98          & 19.4           & 1118         & N/A                   \\
4-R3C4     & HLC Band 3        & circle        & 589.51         & 19.4          & 1236        & 589.51          & 19.4           & 1236         & N/A                   \\
4-R3C5     & HLC Band 4        & circle         & 666.22         & 19.4          & 1397        & 666.22          & 19.4           & 1397         & N/A                   \\
4-R4C1     & Multi-star imaging in Band 4 {[}spare{]}        & square         & 309.07         & 9             & 648         & 309.07          & 9              & 648          & 30                    \\
4-R4C2     & SPC minor-FWHM rotated slit Band 3; short       & rect           & 30.39          & 1             & 64          & 182.32          & 6              & 382          & 15.19                 \\
4-R4C3     & SPC minor-FWHM rotated slit Band 2; tall        & rect           & 27.47          & 1             & 58          & 467.04          & 17             & 979          & 13.74                 \\
4-R4C4     & SPC minor-FWHM rotated slit Band 2; short       & rect           & 27.47          & 1             & 58          & 164.84          & 6              & 346          & 13.74                 \\
4-R4C5     & HLC FWHM vertical slit Band 3                   & rect           & 33.43          & 1.1           & 70          & 182.32          & 6              & 382          & 16.71                 \\
4-R4C6     & SPC major-FWHM vertical slit Band 3; short      & rect          & 60.77          & 2             & 127         & 182.32          & 6              & 382          & 30.39                 \\
4-R5C1     & HLC FWHM curved slit Band 3                     & arc            & 33.43          & 1.1           & 70          & N/A             & $\pm45^\circ$    & N/A          & 16.71                 \\
4-R5C2     & HLC FWHM curved slit Band 3 (reversed)          & arc           & 33.43          & 1.1           & 70          & N/A             & $\pm45^\circ$    & N/A          & 16.71                 \\
4-R5C3     & Left half HLC Band 1                            & arc           & 232.17         & 9.7           & 487         & 464.34          & 19.4           & 974          & 30                    \\
4-R5C4     & Right half HLC Band 1                           & arc            & 232.17         & 9.7           & 487         & 464.34          & 19.4           & 974          & 30                    \\
4-R5C5     & empty / dark                                    & N/A      & 0              & 0             & 0           & 0               & 0              & 0            & N/A                   \\
4-R6C1     & 1.9" radius circular polarimetry field stop     & circle       & 1812.21        & N/A           & 3800        & 1812.21         & N/A            & 3800         & N/A                   \\
4-R6C2     & SPC Wide-FOV Halves Bands 1 and 4                              & puzzle   & 1024.42        & 42.8          & 2148        & 1469.81         & 42.8           & 3082         & 60      \\
4-R6C3     & Long slit \#2                              & rect        & 143.07         & N/A           & 300         & 476.9           & N/A            & 1000         & 30                    \\
4-R6C4     & Long slit \#1                              & rect         & 85.84          & N/A           & 180         & 476.9           & N/A            & 1000         & 30                    \\
4-R6C5     & SPC major-FWHM vertical slit Band 2; tall       & rect          & 54.95          & 2             & 115         & 329.68          & 12             & 691          & 27.47                 \\
4-R6C6     & HLC Band 1 [spare] & circle         & 464.34         & 19.4          & 974         & 464.34          & 19.4           & 974          & N/A  \\
\hline
\end{tabular}
\end{adjustbox}
\caption{Field stops specifications. Mask array positions (starting with 4-) correspond to those shown in Fig.~\ref{fig:fs_array}. Widths and heights will differ slightly from the designed values reported here because of fabrication tolerances as well as observatory and instrument magnification tolerances.
}
\label{tab:fs}
\end{table}

\clearpage

\section{Color Filters}
\label{sec:filters}

All of the color filters included in the Roman Coronagraph are listed in Table \ref{tab:filters}. The four broadband filters, to be used after the dark hole is dug, are named 1F, 2F, 3F, and 4F. (The F stands for full and is generally omitted for brevity.) All the other filters are subsets of the four larger bandpasses and are used for digging the dark holes in separate subbands. There are three subbands each for Bands 1 and 4, and five each for Bands 2 and 3. Because Bands 2 and 3 overlap, though, subbands 3A and 3B are also used to dig the dark hole in Band 2. The narrowest filters, 2C and 3D, are used for spectroscopy wavelength calibration and line spread function measurements, rather than wavefront correction. Note that 2C also coincides with the H$\alpha$ spectral line.

\begin{table}[htbp!]
\centering
\begin{tabular}{c c c }
\hline
Filter Name & Center Wavelength (nm) & FWHM (nm)   \\
\hline
1F          & 575                        & 58              \\
1A          & 555.8                      & 19.2            \\
1B          & 575                        & 19.2           \\
1C          & 594.2                      & 19.2           \\
2F          & 660                        & 112            \\
2A          & 615                        & 22              \\
2B          & 638                        & 18              \\
2C          & 656.3                      & 6.6             \\
3F          & 730                        & 122             \\
3A          & 681                        & 24              \\
3B          & 704                        & 24              \\
3C          & 727                        & 20              \\
3D          & 754                        & 7.5             \\
3E          & 777.5                      & 27              \\
3G          & 752                        & 25              \\
4F          & 825                        & 94              \\
4A          & 792                        & 28              \\
4B          & 825                        & 30              \\
4C          & 857                        & 30              \\
CLEAR       & N/A                        & N/A             \\
ACQ DARK    & N/A                        & N/A            \\
\hline
\end{tabular}
\caption{All color filters and their designed wavelength coverage. The as-built filter properties may differ slightly from these and will be updated in the list of instrument parameters at \url{https://roman.ipac.caltech.edu} when they are available.}
\label{tab:filters}
\end{table}

\acknowledgments 
This work was performed in part at the Jet Propulsion Laboratory, California Institute of Technology, under contract with the National Aeronautics and Space Administration (NASA).

\newpage

\bibliography{bibLibrary_riggs} 
\bibliographystyle{spiebib} 

\end{document}